\documentclass{aa}  

\usepackage{graphicx}
\usepackage{float}
\usepackage{txfonts}
\begin{document}

   \title{Tomography-based observational measurements of the halo mass function via the submillimeter magnification bias}
   \titlerunning{Tomography-based Halo Mass Function measurement with Magnification Bias}
   \authorrunning{Cueli M. M. et al.}

   \author{Cueli M. M.\inst{1,2},
   Bonavera L.\inst{1,2},
    Gonz{\'a}lez-Nuevo J.\inst{1,2},
    Crespo D.\inst{1,2}, Casas J. M.\inst{1,2}, Lapi A.\inst{3,4,5,6}}

  \institute{$^1$Departamento de Fisica, Universidad de Oviedo, C. Federico Garcia Lorca 18, 33007 Oviedo, Spain\\
             $^2$Instituto Universitario de Ciencias y Tecnologías Espaciales de Asturias (ICTEA), C. Independencia 13, 33004 Oviedo, Spain\\
             $^3$SISSA, Via Bonomea 265, 34136 Trieste, Italy\\
    $^4$IFPU - Institute for fundamental physics of the Universe, Via Beirut 2, 34014 Trieste, Italy\\
$^5$IRA-INAF, Via Gobetti 101, 40129 Bologna, Italy\\
$^6$INFN-Sezione di Trieste, via Valerio 2, 34127 Trieste,  Italy}

   \date{}

 
  \abstract
   {}
   {The main goal of this paper is to derive observational constraints on the halo mass fuction (HMF) by performing a tomographic analysis of the magnification bias signal on a sample of background submillimeter galaxies. The results can then be compared with those from a non-tomographic study.}
   {We measure the cross-correlation function between a sample of foreground GAMA galaxies with spectroscopic redshifts in the range $0.1<z<0.8$ (and divided up into four bins) and a sample of background submillimeter galaxies from H-ATLAS with photometric redshifts in the range $1.2<z<4.0$. We model the weak lensing signal within the halo model formalism and carry out a Markov chain Monte Carlo algorithm to obtain the posterior distribution of all HMF parameters, which we assume to follow the Sheth and Tormen (ST) three-parameter and two-parameter fits.}
   {While the observational constraints on the HMF from the non-tomographic analysis are not stringent, there is a remarkable improvement in terms of uncertainty reduction when tomography is adopted. Moreover, with respect to the traditional ST triple of values from numerical simulations, the results from the three-parameter fit predict a higher number density of halos at masses below $\sim10^{12}M_{\odot}/h$ at 95\% credibility. The two-parameter fit yields even more restricting results, with a larger number density of halos below $\sim 10^{13}M_{\odot}/h$ and a lower one above $\sim 10^{14}M_{\odot}/h$, this time at more than 3$\sigma$ credibility. Our results are therefore in disagreement with the standard $N$-body values for the ST fit at $2\sigma$ and $3\sigma$, respectively.}
   {}


   \maketitle
%

\setlength{\abovedisplayskip}{8pt}
\setlength{\belowdisplayskip}{8pt}
\section{Introduction}

The concordance cosmological framework explains the process of galaxy formation  via the hierarchical growth of structure from primordial Gaussian density perturbations. Indeed, virialized dark matter halos break away from the expansion of the background and provide the potential wells for gas to cool down and for galaxies to form. An accurate understanding of the abundance and evolution of dark matter halos is therefore of paramount importance to accurately explain the early Universe and to compute the nonlinear effects of structure formation.

The seminal paper of \citet{ps} laid the conceptual foundations of dark matter halo formation by relating their collapse to the moment that the spherically smoothed linear density field goes over a certain threshold value that depends on cosmology, thus deriving the first analytical model of the halo mass function (HMF). The subsequent work by \cite{bond} established a new analytical approach to dark matter collapse known as the excursion set formalism, which reproduced the results from Press \& Schechter but also provided a much wider framework.

However, given the highly nonlinear character of dark matter halos, N-body simulations have traditionally been the main arena in which to study their formation and abundance. Indeed, a long series of different fits for the HMF has arisen from simulations since the beginning of the century \citep{sheth01,J2001,reed03,reed07,warren06,tinker08,crocce10,courtin11,watson13,despali16,DIE20}, each of them for different mass and redshift ranges. Moreover, baryonic processes such as radiative cooling or feedback from active galactic nuclei are thought to cause non-negligible effects on the abundance of dark matter halos \citep{sawala13,bocquet15,castro20}. Consequently, work toward an observational determination of the abundance of dark matter halos may shed some light on the underlying physics and validate the results from simulations.

With regard to direct measurements of the HMF, it is well known that they present several challenges, such as the Eddington bias of mass-observable relations \citep{ED1913}. Weak lensing-based stacked background shear distributions have thus far been a popular method for calibrating these quantities by means of linking the mean halo mass to observables such as luminosity or stellar mass \citep{SIM17,LEA17,MUR18}. Mass-observable relations can then be exploited to constrain the halo mass function from observations \citep{castro16,li19}.
However, also within the context of gravitational lensing, we propose the angular foreground-background cross-correlation function as an observable of the weak lensing magnification bias effect and a method for shedding light on the direct measurement of the halo mass function. 

Although traditionally considered suboptimal under usual circumstances by comparison with tangential shear measurements, we argue that magnification bias is a promising candidate for cosmological analysis when combined with a sample of background submillimeter galaxies \citep{gonzaleznuevo17,bonavera19,bonavera20}. When a nonzero signal of this cross-correlation is found between a foreground and a background sample of massive objects with nonoverlapping redshift distributions, it signals the physical effect of magnification bias, by which an excess of background sources is measured around foreground sources with respect to what would be found in the absence of the latter \citep{bartelmann01}. Given its direct dependence on the halo mass of the foreground lenses, we propose this observable as a useful observational probe of the HMF under a halo model that only needs additional statistical information about the halo occupation distribution (HOD).

This paper constitutes a first step forward from the recent work of \citet{CUE21}, where the cross-correlation between a foreground sample of GAMA II \citep{DRI11,BAL10,BAL14,LIS15} sources and a foreground sample of H-ATLAS \citep{PIL10,EAL10} submillimeter galaxies was measured with the aim of providing constraints on the halo mass function. By performing a theoretical halo modeling and parametrizing the halo mass function according to the well-known Sheth and Tormen (ST) and Tinker fits, Bayesian posterior probability distributions for their parameters were obtained, and it was concluded that the former model can potentially be well constrained when given enough parameter freedom. This provided a method to obtain tabulated credible intervals of the halo mass function at the redshift of the lenses. However, the present work aims to obtain stronger constraints by performing a tomographic analysis (analogously to \citet{BON21}) that divides foreground sources into four redshift bins in order to incorporate the likely time evolution of the galaxy content of dark matter halos. This approach can also provide information on the universality of the halo mass function by deriving posterior distributions for its parameters at different redshifts. Furthermore, and in light of the work by \citet{GON21}, the cross-correlation function will be measured using two different tiling schemes, the first to keep coherence with \citet{CUE21} and the second because it provides more meaningful statistical conclusions.

The paper has been structured as follows. Section 2 sets the theoretical foundations of this work. We describe the usual framework to characterize the HMF and the parametrization we use as well as the halo model prescription for the angular cross-correlation function. Section 3 details the methodology we follow, namely a description of the data, the measurement process, the different tiling schemes, and the Markov chain Monte Carlo (MCMC) algorithm to sample the posterior distribution. Section 4 presents the results we obtained for both the non-tomographic and tomographic cases and for an additional analysis on universality. Our conclusions are summarized in Section 5.
 
\section{Theoretical framework}

\subsection{The halo mass function}
The (differential) halo mass function $n(M,z)$ is defined as the comoving mean number density of halos at redsfhit $z$ per unit mass. In other words, $\int_{M_1}^{M_2} n(M,z)\,dM$ is the comoving mean number density of halos of mass between $M_1$ and $M_2$ at redshift $z$.

A useful way of parametrizing this quantity is
\begin{equation}
    n(M,z)=\frac{\rho_0}{M^2}f(\nu,z)\,\bigg|\frac{d\ln{\nu(M,z)}}{d\ln{M}}\bigg|\label{HMFnu},
\end{equation}
where 
\begin{equation*}
    \nu(M,z)\equiv\bigg[\frac{\hat{\delta}_c(z)}{\sigma(M,z)}\bigg]^2,
\end{equation*}
$\hat{\delta}_c(z)$ being the linear critical overdensity at redshift $z$ for a region to collapse into a halo at that same redshift and $\rho_0$ the current mean background density. It should be noted that the redshift dependence of $\hat{\delta}_c(z)$ is known to be very weak, but we decided to take it into account through the fit by \cite{kitayama96} given its potential influence on the non-universality of the HMF \citep{courtin11}. Moreover, $\sigma^2(M,z)\equiv D^2(z)\,\sigma^2(M,0)$, where $D(z)$ is the linear growth factor for a $\Lambda$CDM universe (normalized at $z=0$) and $\sigma^2(M,0)$ is the mass variance of the filtered linear overdensity field at present. The function $f(\nu,z)$ depends on the HMF model we consider and, for the purposes of this work, we fix it to the ST fit, that is,
\begin{equation}
    f(\nu,z)=A\sqrt{\frac{a\nu}{2\pi}}\,\bigg[1+\bigg(\frac{1}{a\nu}\bigg)^p\bigg]\,e^{-a\nu/2}.\label{sthmf}
\end{equation}
Several comments should be made about this. Firstly, the choice of the ST fit over other models follows from the results of \cite{CUE21}: for our study, this model seems to compare favorably with other fits, where an important degeneration among parameters is bound to arise together, thus compromising the statistical conclusions, especially when they depend on the prior range. Secondly, the original work by \cite{st1999} imposed a normalization condition accounting for the fact that all mass should be bound up in halos, which allowed them to fix $A$ in terms of $p$ (and resulted in the commonly used best-fit triple $A=0.33$, $a=0.707$ and $p=0.3)$. Following the results from \citet{CUE21}, we opted to leave the $A$ parameter free in most cases, provided however that the condition $\int_0^{\infty} f(\nu)/\nu \,d\nu \le 1$ is met so as to have a meaningful physical interpretation of the HMF. Lastly, it should be noted that the original ST function $f(\nu,z)$ does not depend on redshift, but given the tomographic character of this work, we aim to study the possible time evolution of its parameters to quantify deviations from universality.

\subsection{The foreground-background cross-correlation function}
Aside from galaxy tangential shear measurements, the galaxy-mass correlation can be probed via the foreground-background cross-correlation function in the context of weak gravitational lensing. In the case of nonoverlapping redshift distributions, this quantity can be expressed as \citep{cooray02}
\begin{align}
    w_{fb}(\theta)&=2(\beta-1)\int_0^{\infty}\frac{dz}{\chi^2(z)}n_f(z)W^{\text{lens}}(z)·\nonumber\\
    &\int_0^{\infty}dl\frac{l}{2\pi}P_{\text{g-dm}}(l/\chi(z),z)J_0(l\theta)\label{crosscorr},
\end{align}
where
\begin{equation*}
    W^{\text{lens}}(z)=\frac{3}{2}\frac{H_0^2}{c^2}\bigg[\frac{E(z)}{1+z}\bigg]^2\int_z^{\infty}dz'\frac{\chi(z)\chi(z'-z)}{\chi(z')}n_b(z'),
\end{equation*}
$n_b(z)$ ($n_f(z)$) being the unit-normalized background (foreground) source distribution, $\chi(z)$ the comoving distance at redshift $z$, $J_0$ the zeroth-order Bessel function of the first kind, and $\beta$ the logarithmic slope of the background source number counts, commonly fixed to 3 for submillimeter galaxies \citep{LAP11,CAI13,BIA15,BIA16,gonzaleznuevo17,bonavera19,bonavera20}.

The nonlinear galaxy-dark matter cross-power spectrum is parametrized according to the halo model formalism as described in detail in Appendix A. Besides cosmology, it depends on the halo density profile, which we assume to match an Navarro-Frenk-White (NFW) model \citep{navarro97}, and the HOD parameters, which describe how galaxies populate halos on average. Indeed, according to the three-parameter model by \cite{zehavi05}, the mean number of galaxies within a halo of mass $M$ is given by 
\begin{equation}
    \langle N_g\rangle_M=\langle N_{c_g}\rangle_M+\langle N_{c_s}\rangle_M=\bigg[1+\bigg(\frac{M}{M_1}\bigg)^{\alpha}\bigg]\,\Theta(M-M_{min})\label{Ngalaxies},
\end{equation}
where $\langle N_{c_g}\rangle_M$ and $\langle N_{s_g}\rangle_M$ are the mean number of central and satellite galaxies in a halo of mass $M$, respectively. Parameter $M_{\text{min}}$ describes the minimum mean mass for a halo to host a (central) galaxy, while $M_1$ represents the mean halo mass at which there is exactly one satellite galaxy. As described in Appendix A, we define halos as spherical overdense regions with an overdensity equal to the virial value, as estimated via the fit by \cite{weinberg03} and a concentration parameter given by that of \cite{bullock01}.

As a consequence, for a fixed cosmology, which we assume to match \emph{Planck}'s \citep{planck18VI}, the angular cross-correlation function \eqref{crosscorr} depends on both the HMF and the HOD parameters. Our goal is thus to derive posterior probability distributions for these two sets of parameters.

\section{Work methodology}

\subsection{Galaxy samples}
The foreground and background sources have been extracted, respectively, from the GAMA II \citep{DRI11,BAL10,BAL14} and H-ATLAS \citep{PIL10,EAL10} surveys, which were coordinated to maximize the common area. Indeed, they both covered the three equatorial regions at 9, 12 and 14.5 h and part of the south galactic pole, amounting to a common area of $\sim207\text{deg}^2$. 

\begin{figure}[h]
    \centering
    \includegraphics[width=\columnwidth]{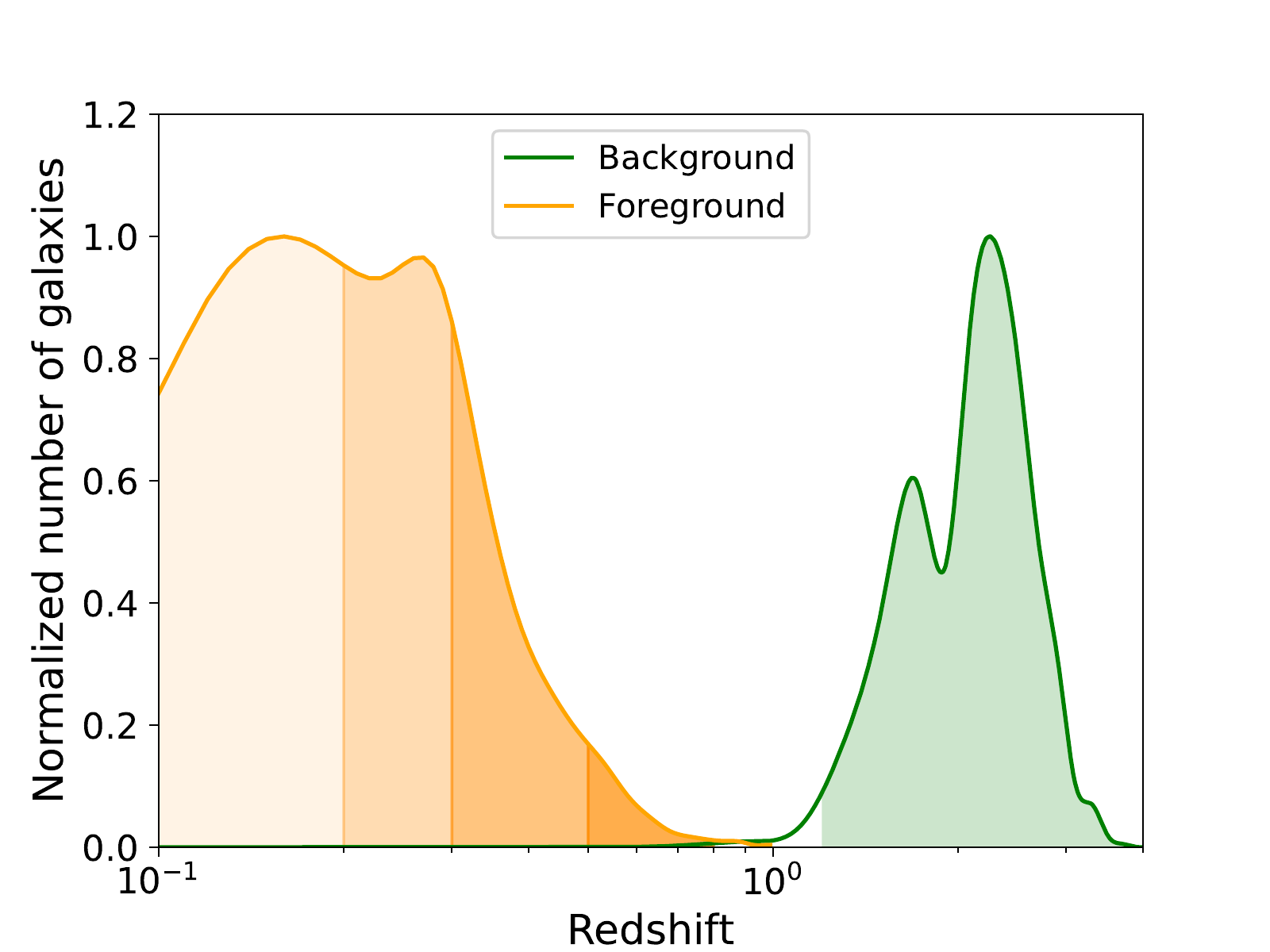}
    \caption{Normalized redshift distribution of the background (in green) and foreground (in orange) samples of galaxies. The shaded regions represent the selected redshift ranges: $1.2<z<4.0$ for the background (in green) and $0.1<z<0.2$, $0.2<z<0.3$, $0.3<z<0.5$ and $0.5<z<0.8$ for the four different foreground bins (in orange).}
    \label{z_distributions}
\end{figure}

The foreground sample consists, in principle, of GAMA II sources within the interval $0.1<z<0.8$, which results in $\sim$ 225000 galaxies with a median spectroscopic redshift of $z_{\text{med}}=0.28$. This is the same sample that was used in \citet{CUE21}, but given the tomographic character of this work, it has been further divided into four redshift bins, namely, $[0.1,0.2]$, $[0.2,0.3]$, $[0.3,0.5]$ and $[0.5,0.8]$, referred to as bin 1 to bin 4. It should be noted that this particular choice was made to ensure a similar number of sources in each bin. Moreover, we have included additional sources with respect to our previous works \citep{gonzaleznuevo17,bonavera20,CUE21}, which excluded galaxies with $z<0.2$. The redshift distribution of our foreground sample is shown in Fig. \ref{z_distributions} (in orange), where the four different bins are also depicted in different color shades. Given the spectroscopic nature of the measured redshifts, no source is assigned the wrong bin.

The background sample has been selected from the H-ATLAS sources detected in the aforementioned regions. In order to guarantee that there was no overlap with the foreground sample, a photometric redshift selection in the interval $1.2<z<4.0$ has been applied, which results in $\sim$ 57930 galaxies. The procedure of redshift estimation is extensively described in \cite{gonzaleznuevo17} and \cite{bonavera19}, but mainly consists in a $\chi^2$ fit to the SPIRE (and PACS when available) data of the SMM J2135-0102 template spectral energy distribution \citep["the Cosmic Eyelash";][]{IVI10,SWI10}, which was found to be the best overall fit. The resulting redshift distribution of our background sample if shown in Fig. \ref{z_distributions} (in green), where the random errors in photometric redshift estimation have been taken into account, as explained in \cite{gonzaleznuevo17}.

\begin{figure}[h]
    \centering
    \includegraphics[width=\columnwidth]{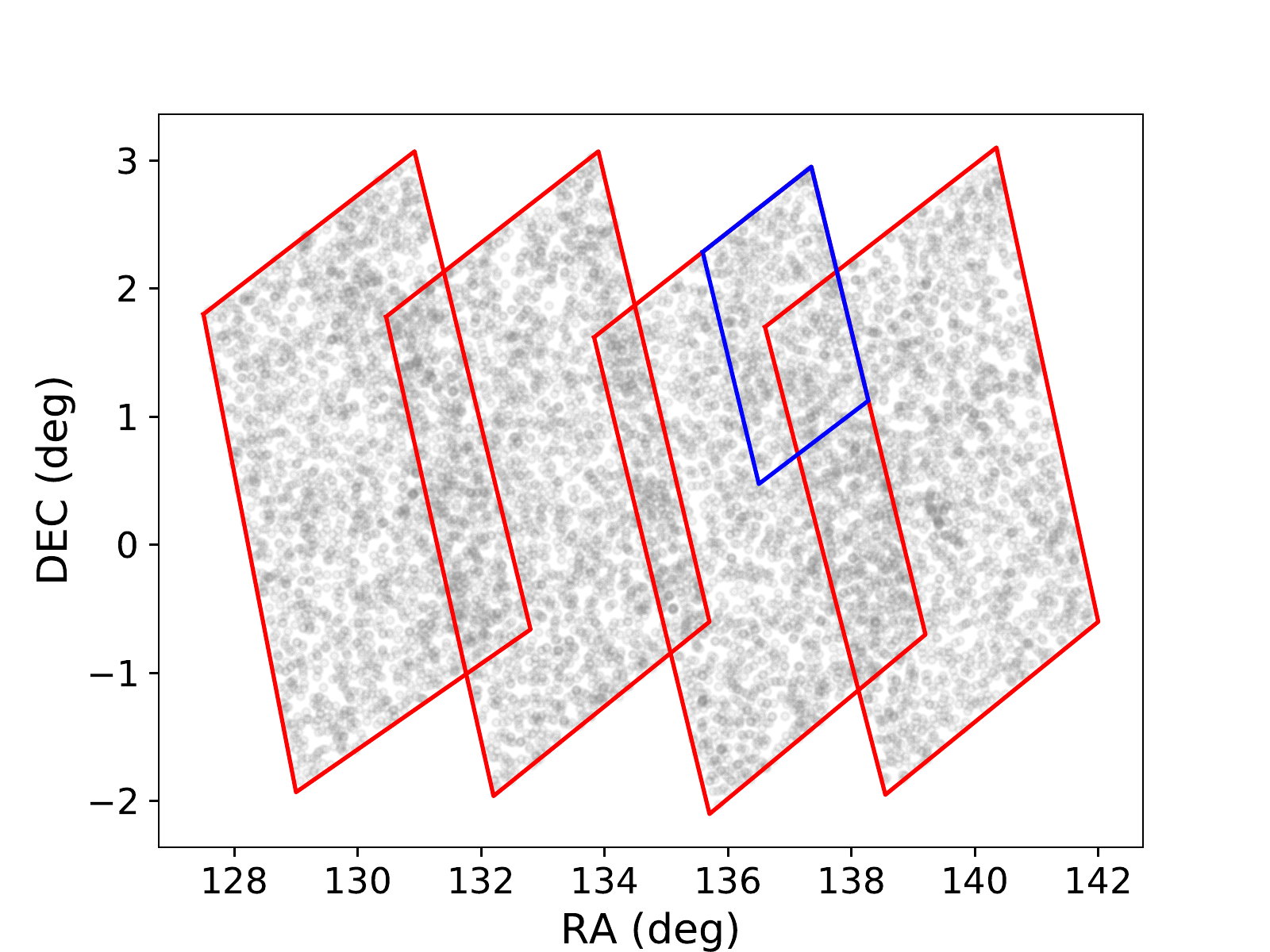}
    \caption{Representation of the tiles (in red) and minitiles (in blue) schemes for the 9h GAMA field. The pattern is analogous for the rest.}
    \label{tiles_minitiles_scheme}
\end{figure}

\begin{figure*}[t]
    \centering
    \includegraphics[width=0.8\textwidth]{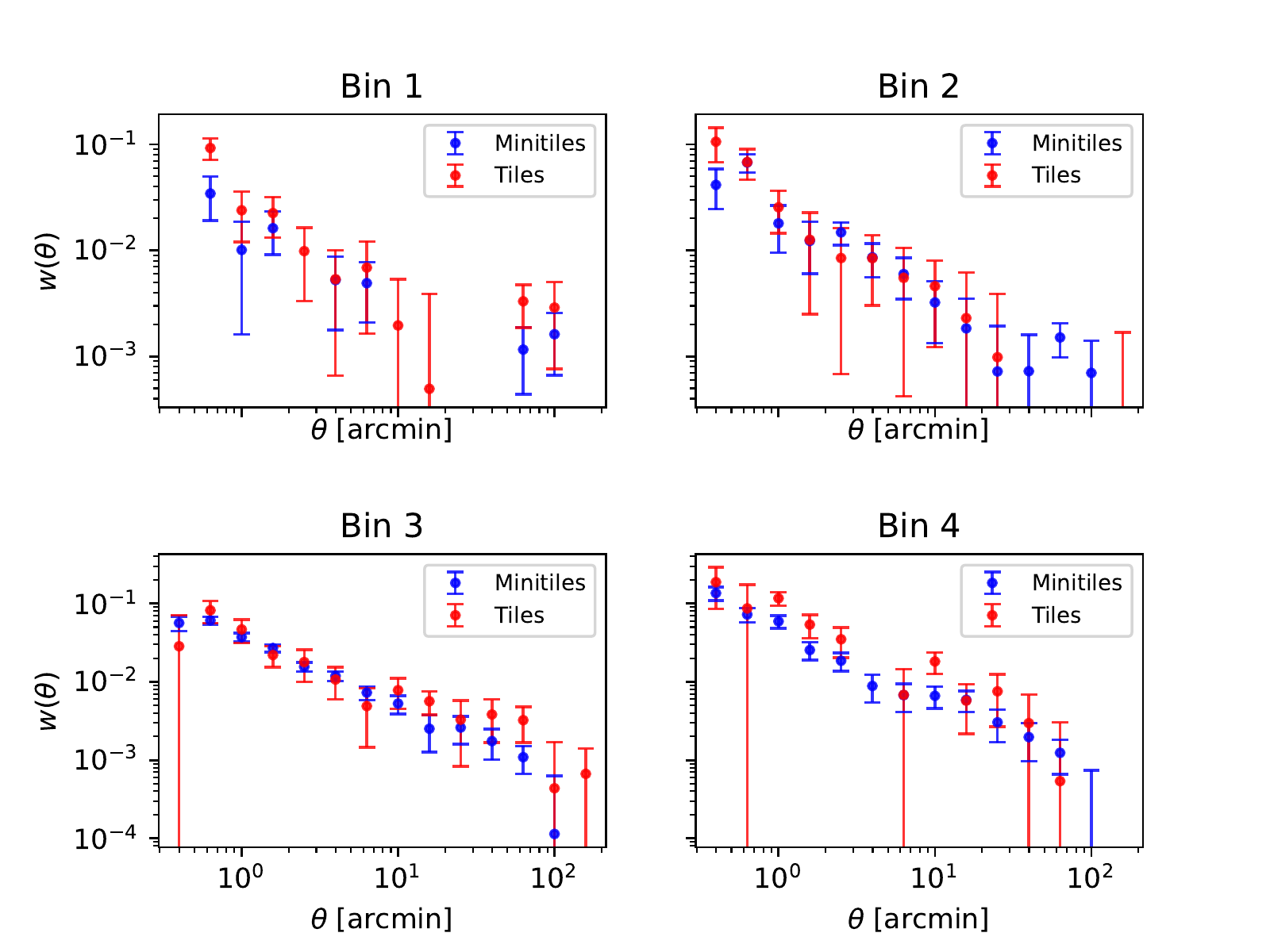}
    \caption{Cross-correlation measurements at different $\theta$ values for the tiles (in red) and minitiles (in blue) schemes. The data for foreground bins 1 to 4 are depicted from left to right and from top to bottom. }
    \label{xc_data_tiles_and_minitiles}
\end{figure*}

\subsection{Measurement of the cross-correlation function}

The H-ATLAS scanning strategy produced characteristic overlapping rhomboidal shapes or "tiles" in most of the fields. Considering the area of study, we have \text{16} such tiles, each with an area of $\sim$ 16 $\text{deg}^2$. As thoroughly described in \cite{GON21}, the choice of different tiling schemes to measure the cross-correlation function (for example, dividing each tile into four "minitiles" in order to have more regions over which to average; see Fig. \ref{tiles_minitiles_scheme}) brings along different large-scale biases in the measurements that need to be corrected so as to have a consistent method to estimate our observable. In principle, since the present work aims to extend that of \citet{CUE21}, where the tile scheme was adopted, coherence would require the same approach be used. However, in view of the work by \cite{GON21}, we opt to also consider the minitile scheme at first given the minimum surface area corrections that are required and, most importantly, the larger sample size, as will now be emphasized.

For each tiling scheme (tiles or minitiles) and redshift bin, we estimated the angular cross-correlation function via a modified version of the \cite{landy93} estimator \citep{herranz01},
\begin{equation*}
    \tilde{w}^i(\theta)=\frac{\text{D}_{\text{f}}^i \text{D}_{\text{b}}(\theta)-\text{D}_{\text{f}}^i \text{R}_{\text{b}}(\theta)-\text{D}_{\text{b}}
    \text{R}_{\text{f}}^i(\theta)+\text{R}_{\text{f}}^i \text{R}_{\text{b}}(\theta)}{\text{R}_{\text{f}}^i \text{R}_{\text{b}}(\theta)},
\end{equation*}
where the superscript $i$ denotes the $i_{\text{th}}$ redshift bin and $\text{D}_{\text{f}}\text{D}_{\text{b}}(\theta)$, $\text{D}_{\text{f}}\text{R}_{\text{b}}(\theta)$, $\text{D}_{\text{b}}\text{R}_{\text{f}}(\theta)$ and $\text{R}_{\text{f}}\text{R}_{\text{b}}(\theta)$ denote the normalized foreground-background, foreground-random, background-random and random-random pair counts for the angular separation $\theta$. The random catalog for each tile or minitile is generated from mock random positions for ten times the number of lenses and sources, respectively. As studied in detail by \citet{GON21}, due to the relatively low surface density of the foreground sample and the size of the tiles or minitiles, there is no need of a more complex building procedure because the surface density variations can be considered negligible.
In addition, the cross-correlation measurement for each tile or minitile is obtained by averaging over ten different random (Poisson) catalog realizations. In order to minimize cosmic variance, the final value for a given $\theta$ corresponds to the mean over tiles or minitiles, with an associated uncertainty of $\sigma/\!\sqrt{n}$, where $\sigma$ is the standard deviation of the average measurements and $n$ is the number of tiles or minitiles. Finally, to take into account the finite area of the tiles/minitiles, the integral constraints estimated by \citet{GON21}, $5\cdot10^{-4}$ and $9\cdot10^{-4}$ respectively, are used to correct the estimated cross-correlation functions. This correction is, however, mainly relevant at the largest scales ($>20$ arcmin).

The data obtained for each foreground redshift bin are shown in Fig. \ref{xc_data_tiles_and_minitiles}, where the red dots refer to the tiles scheme and the blue to the minitiles scheme. It should be noted that cross-correlation measurements performed with the tiles scheme generally have larger error bars. This is the expected behavior, given the fact that the number of regions over which to average is about four times as big in the minitiles case. In fact, some measurements at smaller scales show significant differences between both tiling schemes that were not present in the non-tomographic analysis by \cite{GON21}. This is most likely due to the reduction in sample size due to splitting the foreground galaxies in several bins.

It should be noted that, on top of the measurements for each individual redshift bin, $\tilde{w}^i(\theta)$, we performed (with the same approach) a single-bin computation of the cross-correlation function, $\tilde{w}(\theta)$, in order to compare the tomographic results with those of a non-tomographic analysis in the redshift bin $[0.1-0.8]$.

\subsection{Parameter estimation}
In order to derive Bayesian posterior probability distributions for the involved parameters, we performed an MCMC analysis using the open source \emph{emcee} software package \citep{foreman13}, a Python-based implementation of the Goodman \& Weare affine invariant MCMC ensemble sampler \citep{goodman10}. An important comment should be made before going further: \emph{emcee} is not an optimizer, but a sampler. As such, the maximum of the $N$-dimensional posterior distribution (that is, the unique tuple of parameters that best fits the data) we obtain from it is not a reliable
estimate of the parameters, since there can be a considerable $N$-dimensional degeneracy. Due to this degeneracy, other tuples with relatively different values can provide as good a fit\footnote{On top of this, $N$-dimensional credible regions are neither feasible to compute nor useful.}. This single tuple could be plotted to ensure the sampling was correctly carried out, but the full statistical picture is provided by the $N$-dimensional posterior distribution. When one derives the marginalized distribution for a single parameter, its mode, mean or median need not coincide with the corresponding value from the $N$-dimensional best fit (it can even result in a poor fit!). The bottom line is that point estimates should be treated with care. 

We carried out different MCMC runs assuming that errors are Gaussian and independent and parametrizing the HMF with the ST fit, following the conclusions of \citet{CUE21}. It is at this very point that the choice of a tiling scheme becomes relevant. If the sample size for a given $\theta$ value is not large enough, statistical conclusions may be compromised since the random variable $w(\theta)$ may significantly deviate from pure Gaussianity. As a consequence, both schemes will be analyzed in the non-tomographic case, since relevant differences are not expected, but only the results coming from averaging over minitiles will be further taken into account, since these provide the most rigorous statistical conclusions. 

Firstly, we performed a non-tomographic (or single-bin) analysis with both tiling schemes. If $\{P_i\}_i$ denotes the set of parameters to be estimated in the run, the log-likelihood function is expressed as
\begin{equation*}
\begin{split}
    \log \mathcal{L}(\theta_1,\ldots,\theta_m;\{P_i\}_i)=-\frac{1}{2}&\sum_{j=1}^{m}\Big[\log{2\pi\sigma_j^2}+\\
    &+\frac{[w(\theta_j;\{P_i\}_i)-\tilde{w}(\theta_j))]^2}{\sigma_j^2}\Big],
\end{split}
\end{equation*}
where $\sigma_j$ is the error in the $j_{\text{th}}$ measurement and $w(\theta_j)$ and $\tilde{w}(\theta_j)$ are the theoretical and measured value of the cross-correlation function at angular scale $\theta_j$. Moreover, $\{P_i\}_i=\{A,a,p,\alpha,M_{min},M_1\}$.

Secondly, we performed a tomographic run in all four redshift bins altogether, that is, we considered the joint likelihood for all bins. To analyze the effect that a time evolution of the HOD could have in the universal ST fit, the HMF parameters are not considered to vary among bins. The log-likelihood function thus reads\footnote{The cross-correlation is not measured at exactly the same $\theta$ values in all bins, as could appear from the equation, but we have chosen to write it in this way so as not to overcomplicate the expression.}
\begin{equation*}
\begin{split}
    \log{\mathcal{L}(\theta_1,\ldots,\theta_m;\{P_i\}_i})=-\frac{1}{2}&\sum_{k=1}^{4}\sum_{j=1}^{m}\Big[\log{2\pi(\sigma_j^k)^2}+\\
    &+\frac{[w^k(\theta_j;\{P_i\}_i)-\tilde{w}^k(\theta_j))]^2}{(\sigma_j^k)^2}\Big],
\end{split}
\end{equation*}
where $\sigma^k_j$ is the error in the $j_{\text{th}}$ measurement in redshift bin $k$ and $w_k(\theta_j)$ and $\tilde{w}_k(\theta_j)$ are the theoretical and measured value of the cross-correlation function in redshift bin $k$ at angular scale $\theta_j$. Moreover, $\{P_i\}_i=\{A,a,p,\alpha_1,M_{min_1},M_{1_1},\alpha_2,M_{min_2},M_{1_2},\alpha_3,$\\$M_{min_3},M_{1_3},\alpha_4,M_{min_4},M_{1_4}\}$.
It should be noted that the assumption of statistical independence of the cross-correlation measurement among different bins is justified: there is no possible bin overlap in redshift as the foreground redshifts are all spectroscopic and the probability of double lensing events or of an alignment of two lenses from different bins and a background galaxy is very low. Considering the typical lensing probability of order $10^{-3}$, the double lensing probability would be of order $10^{-6}$, thus completely negligible from the statistical point of view \citep[see][and references therein]{BON21}.

Lastly, we performed individual runs in each redshift bin separately in order to account for the possible redshift variation of the HMF parameters. The log-likelihood function for redshift bin $k$ is then expressed as

\begin{equation*}
\begin{split}
    \log{\mathcal{L}^k(\theta_1,\ldots,\theta_m;\{P^k_i\}_i})=-\frac{1}{2}&\sum_{j=1}^{m}\Big[\log{2\pi(\sigma_j^k)^2}+\\
    &+\frac{[w^k(\theta_j;\{P^k_i\}_i)-\tilde{w}^k(\theta_j))]^2}{(\sigma_j^k)^2}\Big],
\end{split}
\end{equation*}
where $\{P^k_i\}=\{A_k,a_k,p_k,\alpha_k,M_{min_k},M_{1_k}\}$.

With regard to the prior probability distributions, they were chosen to be uniform with a wide enough range for all the involved parameters to find a reasonable balance between informational bias and computation efficiency, as shown in the tables throughout the paper. The delicate issue regarding the choice of the allowed portion of parameter space in an MCMC analysis was raised in \cite{CUE21} and the discussion therein led to the conclusion that the $p$ parameter in the ST fit should be allowed to take nonpositive values. 

Lastly, the full posterior distribution of the set of parameters in each case is sampled to derive observational constraints on the HMF for several mass values at any lens redshift. Indeed, although marginalized distributions are also obtained for all parameters, it is the $N$-dimensional probability distribution as a whole (which in our case will be summarized by a median value and credible intervals for the HMF at different halo masses) that provides the full picture and, not surprisingly, that proves useful for our purposes.

\begin{table*}[t]
\caption{Parameter prior distributions and statistical results for the non-tomographic run using the tiles scheme.}
\centering
\begin{tabular}{c c c c c c c}
\hline
\hline
Parameter & Prior & Median & Mean & $68\%$ CI& $95\%$ CI\\
\hline 
$A$&$\mathcal{U}$[0,1]&$0.18$&$0.21$&$[0.03,0.28]$&$[0.00,0.46]$\\
$a$&$\mathcal{U}$[0,10]&$3.37$&$3.93$&$[0.85,5.16]$&$[0.95.10.00]$\\
$p$&$\mathcal{U}$$[-10,0.5]$&$-0.83$&$-0.94$&$[-1.48,0.19]$&$[-2.54,0.50]$\\
$\alpha$&$\mathcal{U}$$[0,1.5]$&$0.88$&$0.83$&$[0.60,1.50]$&$[0.00,1.50]$\\
$\text{log}{M}_{min}$&$\mathcal{U}$$[9,16]$&$12.51$&$12.18$&$[11.96,13.09]$&$[9.94,13.35]$\\
$\text{log}{M}_{1}$&$\mathcal{U}$[9,16]&$11.77$&$12.07$&$[9.00,13.18]$&$[9.00,16.00]$\\
\hline
\hline
\end{tabular}
\label{table_nontomo_tiles}
\tablefoot{The columns denote the parameter in question, its prior distribution and the mean, median, and $68\%$ and $95\%$ credible intervals of its marginalized one-dimensional posterior distribution. Parameters $M_{\text{min}}$ and $M_1$ are expressed in $M_{\odot}/h$.}
\end{table*}

\begin{table*}[h]
\caption{Parameter prior distributions and statistical results for the non-tomographic run using the minitiles scheme.}
\centering
\begin{tabular}{c c c c c c c}
\hline
\hline
Parameter & Prior & Median & Mean & $68\%$ CI& $95\%$ CI\\
\hline 
$A$&$\mathcal{U}$[0,1]&$0.16$&$0.18$&$[0.02,0.24]$&$[0.00,0.42]$\\
$a$&$\mathcal{U}$[0,10]&$2.85$&$3.34$&$[0.68,4.28]$&$[0.00,8.00]$\\
$p$&$\mathcal{U}$$[-10,0.5]$&$-0.99$&$-1.09$&$[-1.74,0.14]$&$[-2.78,0.50]$\\
$\alpha$&$\mathcal{U}$$[0,1.5]$&$0.93$&$0.88$&$[0.68,1.50]$&$[0.00,1.50]$\\
$\text{log}{M}_{min}$&$\mathcal{U}$$[9,16]$&$12.35$&$12.02$&$[11.72,13.03]$&$[9.76,13.27]$\\
$\text{log}{M}_{1}$&$\mathcal{U}$[9,16]&$10.90$&$11.17$&$[9.00,11.72]$&$[9.00,14.26]$\\
\hline
\hline
\end{tabular}
\label{table_nontomo_minitiles}
\tablefoot{The column information is the same as Table \ref{table_nontomo_tiles}}
\end{table*}

\section{Results}

\subsection{Non-tomographic analysis}

\begin{figure}[h]
    \centering
    \includegraphics[width=0.9\columnwidth]{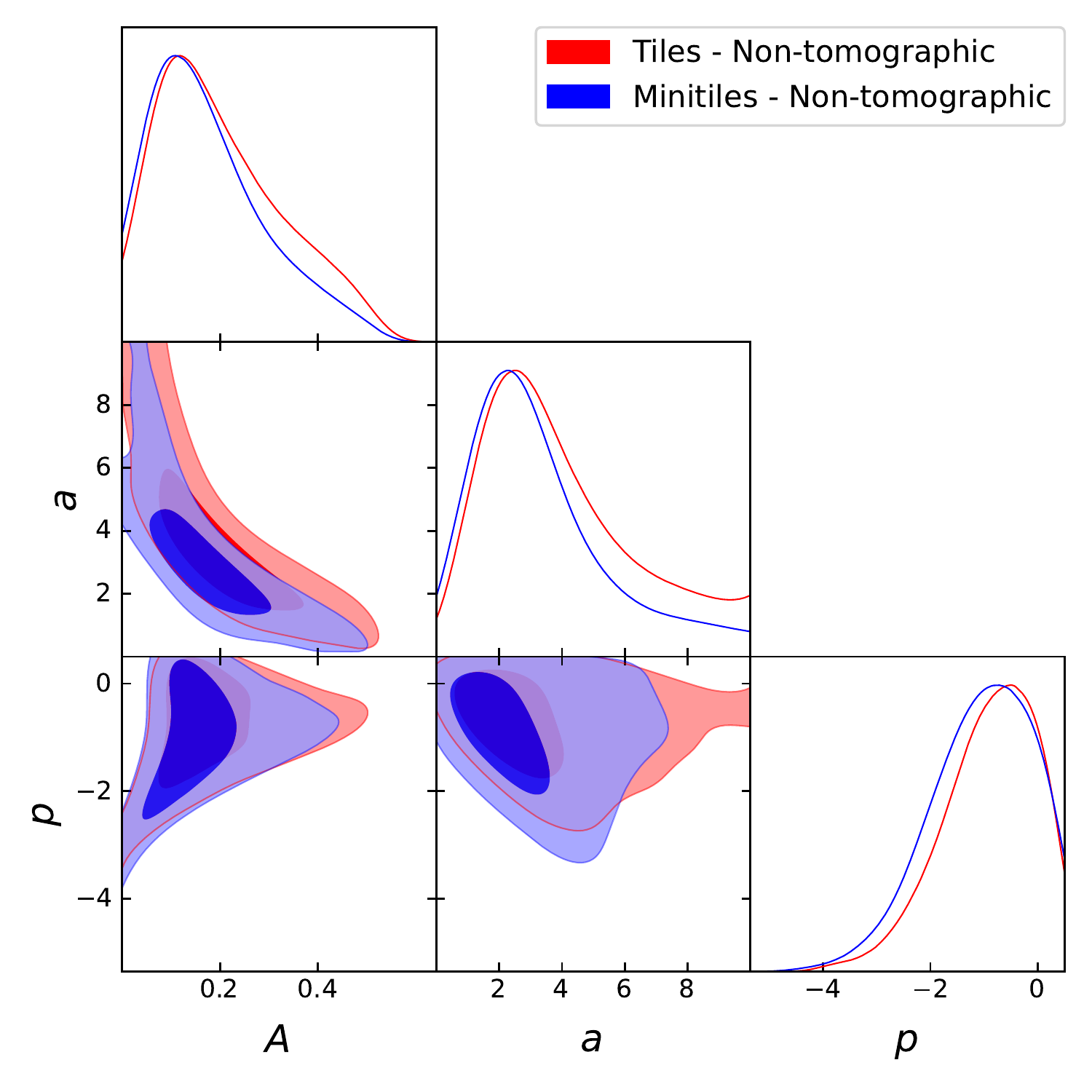}
    \caption{One- and two-dimensional (contour) posterior distributions for the HMF parameters from the non-tomographic run with the tiles (in red) and minitiles (in blue) schemes.}
    \label{hmfcorner_nontomo_tiles_vs_minitiles}
\end{figure}

\begin{figure*}[t]
\centering
\minipage{0.32\textwidth}
  \includegraphics[width=\linewidth]{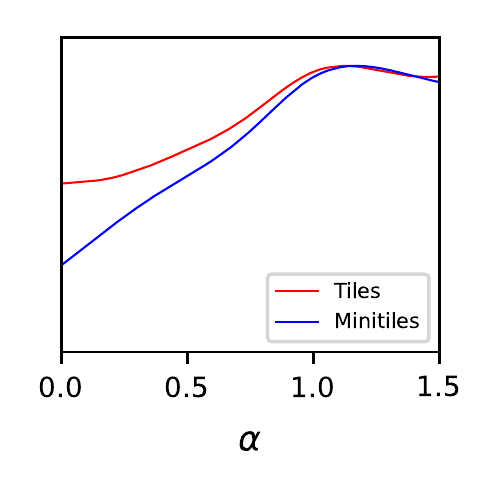}
\endminipage\hfill
\minipage{0.32\textwidth}
  \includegraphics[width=\linewidth]{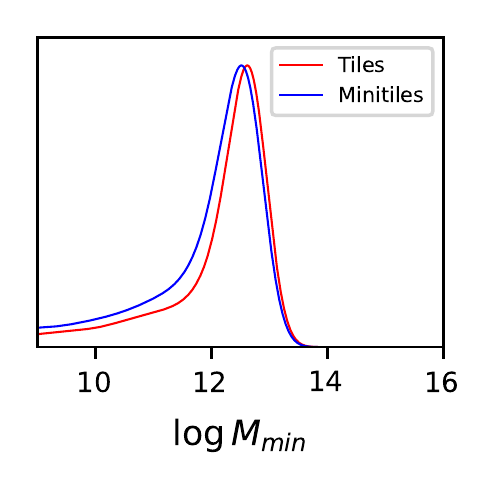}
\endminipage\hfill
\minipage{0.32\textwidth}%
  \includegraphics[width=\linewidth]{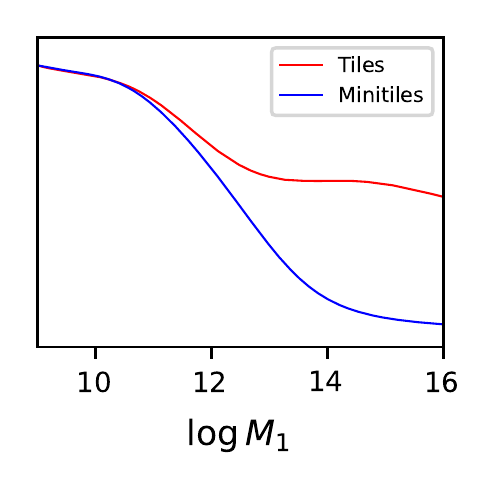}
\endminipage
\caption{One-dimensional posterior distributions for the HOD parameters from the non-tomographic run using the tiles (in red) and minitiles (in blue) scheme.}
\label{1D_nontomo_tiles_and_minitiles}
\end{figure*}

Table \ref{table_nontomo_tiles} shows the summarized statistical results from the non-tomographic MCMC run with the tiles scheme. For visual purposes, the part of the full corner plot showing only the HMF parameters is depicted in Fig. \ref{hmfcorner_nontomo_tiles_vs_minitiles} (in red) and the corresponding marginalized posterior distribution of the HOD parameters is shown in Fig. \ref{1D_nontomo_tiles_and_minitiles} (in red). Figure \ref{corner_nontomo_full_tiles_vs_minitiles} (in red) depicts the full corner plot.

As can be seen in Fig. \ref{hmfcorner_nontomo_tiles_vs_minitiles} (in red), all three HMF marginalized distributions display clear peaks and confirm the findings from \cite{CUE21} with fixed HOD values: the submillimeter galaxy magnification bias constrains the ST parameters when $p$ is allowed to take negative values and $A$ is left free. 

As expected, the uncertainties increase with respect to the fixed HOD case, but there is little to no variation in the $p$ parameter and only a widening of the curve for $a$ (the mode remaining practically unchanged). As for the normalization parameter, $A$, its distribution is displaced to smaller values because the freedom in the HOD allows it to; for instance, as shown by Fig. \ref{corner_nontomo_full_tiles_vs_minitiles}, fixed values of 12.4 and 13.6 for $\log{M}_{\text{min}}$ and $\log{M}_1$ imply lower $A$ values. More precisely, the mean and 68\% credible intervals are $A=0.21^{+0.07}_{-0.18}$, $a=3.93^{+1.23}_{-3.08}$, and $p=-0.94^{+1.13}_{-1.54}$, with median values of 0.18, 3.37, and -0.83, respectively. It should be noted that, as in the analysis of \citet{CUE21} with fixed HOD parameters, $p$ clearly prefers negative values as opposed to the standard 0.3 from $N$-body simulations. Moreover, and since the $a$ distribution sources the majority of the uncertainty in the HMF, these results already hint not very restrictive constraints on this quantity, as will be revealed later.

\begin{figure}[h]
    \centering
    \includegraphics[width=\columnwidth,height=7.5cm]{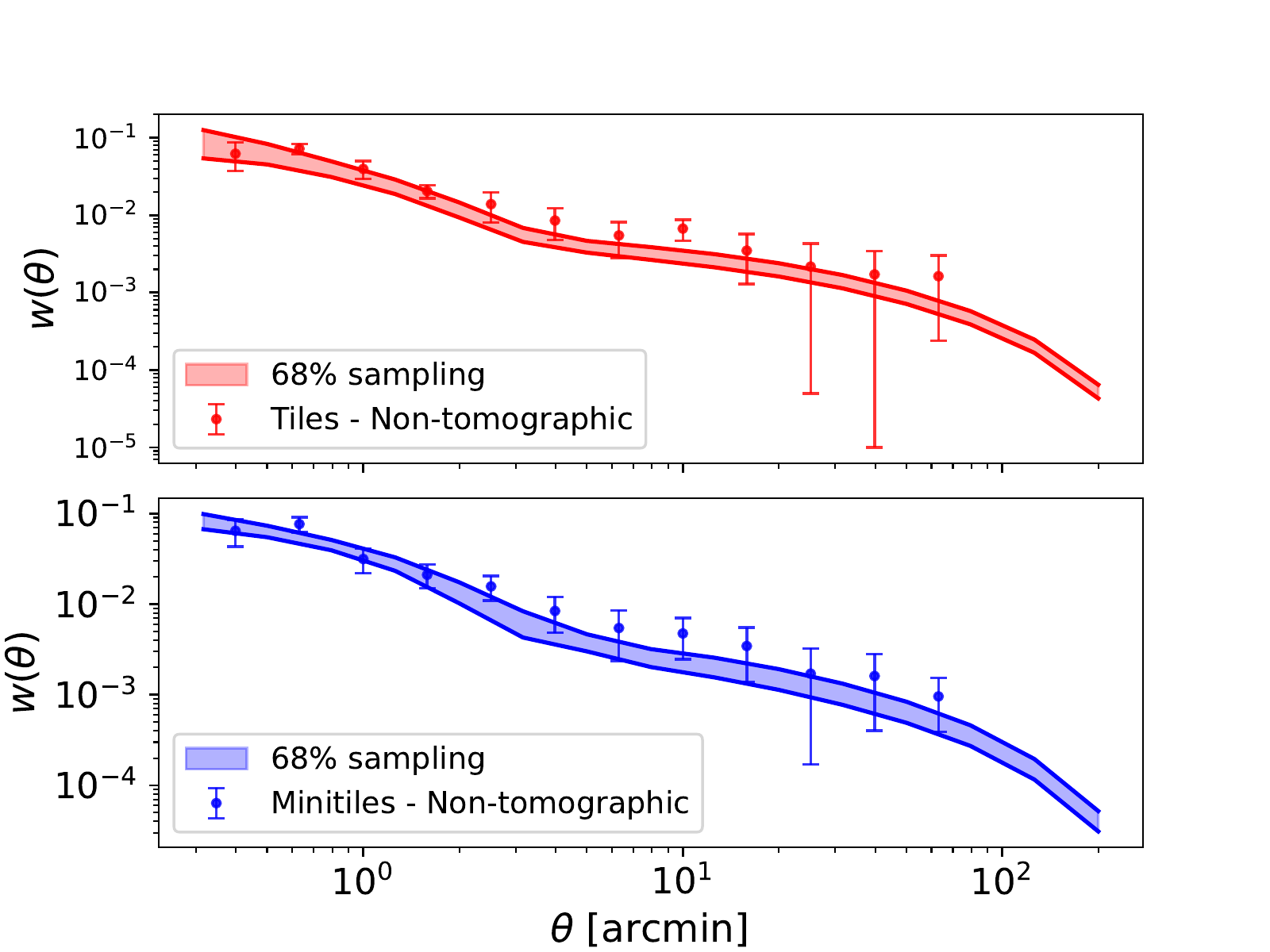}
    \caption{Posterior-sampled cross-correlation function (68\% band) and measured data for the non-tomographic case with the tiles scheme (top panel) and the minitiles scheme (bottom panel).}
    \label{xc_sampled_nontomo_tiles_vs_minitiles}
\end{figure}

As for the marginalized distribution of the HOD parameters, Fig. \ref{1D_nontomo_tiles_and_minitiles} shows a well-constrained Gaussian-like posterior for the minimum mean halo mass, $\log{M}_{\text{min}}=12.18^{+0.91}_{-0.22}$ (with a median of $12.51$, whereas low values for $M_1$ seem to be slightly preferred without ruling out the highest masses due to the restriction that $M_{min}$ should be less than $M_1$: $\log{M}_1=11.17^{+0.55}_{-2.17}$ with a median of 10.90. As already expected given the results from previous works, the $\alpha$ parameter is unconstrained, given its little influence on our observable. The 68\% full posterior sampling of the cross-correlation is shown in Fig. \ref{xc_sampled_nontomo_tiles_vs_minitiles} (top panel), along with the data corresponding to the tiles scheme, confirming that the MCMC algorithm was correctly carried out.

Table \ref{table_nontomo_minitiles} shows the corresponding statistical results for the non-tomographic run with the minitiles scheme. As in the previous case, the part of the full corner plot showing only the HMF parameters and the marginalized posterior distributions of the HOD parameters are depicted in Fig. \ref{hmfcorner_nontomo_tiles_vs_minitiles} (in blue) and in Fig. \ref{1D_nontomo_tiles_and_minitiles} (in blue), respectively. Figure \ref{corner_nontomo_full_tiles_vs_minitiles} (in blue) depicts the full corner plot.

From a qualitative point of view, Figs. \ref{hmfcorner_nontomo_tiles_vs_minitiles} and \ref{1D_nontomo_tiles_and_minitiles} show a similar behavior of all six parameters with respect to the previous case using the tiles scheme. However, as clearly revealed by Fig. \ref{corner_nontomo_full_tiles_vs_minitiles}, there are some slight changes. As regards the HMF parameters, all three marginalized distributions are somewhat displaced toward lower values and those of $A$ and $a$ are narrower; indeed, the mean and 68\% credible intervals are $A=0.18^{+0.06}_{0.16}$, $a=3.34^{+0.94}_{-2.66}$ and $p=-1.09^{+1.23}_{-0.65}$, with median values  of 0.16, 2.85 and -0.99, respectively.

With respect to the HOD parameters, and in comparison to the previous case, the main difference is that higher values of $M_1$ are more strongly suppressed, with an 68$\%$ and 95\% upper bound of 11.72 and 14.26 for $\log{M_1}$. As for $M_{\text{min}}$, it barely changes with respect to using the tiles scheme, with a mean of $\log{M_{\text{min}}}=12.02^{+1.01}_{-0.30}$
The $\alpha$ parameter is again unconstrained, although can be assigned a highest posterior density 68\% lower bound of 0.68. The corresponding 68\% posterior sampling of the cross-correlation function is shown in Fig. \ref{xc_sampled_nontomo_tiles_vs_minitiles} (bottom panel), along with the data corresponding to the minitiles scheme, confirming once more that the MCMC algorithm was correctly carried out.

\begin{figure}[h]
    \centering
    \includegraphics[width=0.85\columnwidth]{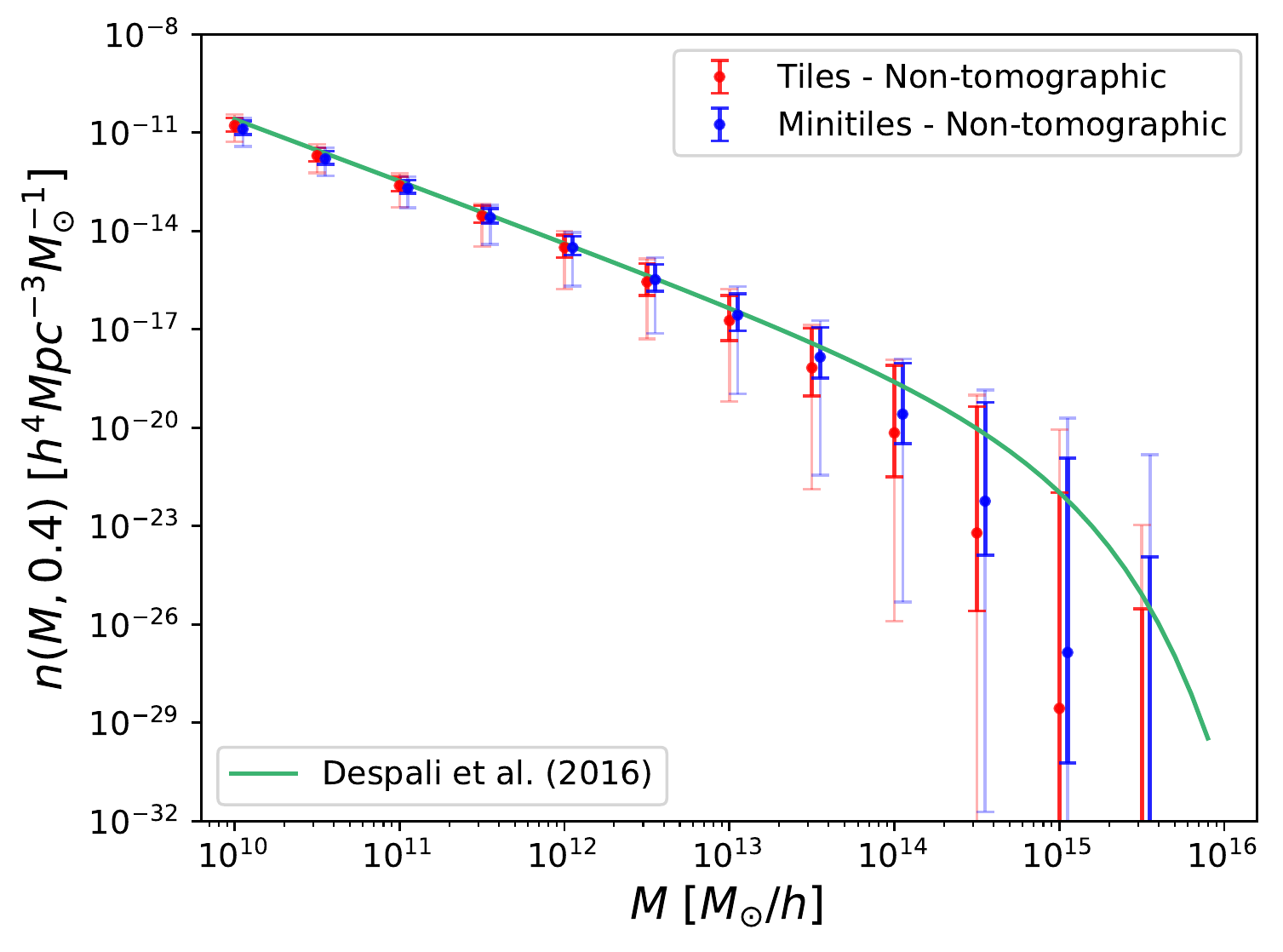}
    \caption{Credible intervals (68\% in bold and 95\% in faint colors) for the $z=0.4$ HMF at different mass values when the full posterior distribution is sampled in the non-tomographic setup within the minitiles (blue) and tiles (red) schemes. The ST best fit by \cite{despali16} is drawn with a green line.}
    \label{tabulated_hmf_no_tomo}
\end{figure}

Moreover, and most importantly for the purpose of this paper (the observational constraining of the number density of halos), credible intervals for the HMF at any lens redshift can be obtained via the sampling of the full posterior distribution. Fig. \ref{tabulated_hmf_no_tomo} shows the median values and the 68\% and 95\% credible intervals at several masses for $z=0.4$ in the non-tomographic case for both the tiles and minitiles scheme (in red and blue, respectively), along with the ST best fit found by \cite{despali16} under a \emph{Planck} cosmology (green line). One should keep in mind that the algorithm used by \cite{st1999} identified halos as spherically overdense regions with an overdensity equal to 178 times the mean density; our mass definition is thus different since it relies on the virial value, and this is why direct comparisons should be performed under the same halo mass definition, hence the aforementioned choice. While in agreement with the usual values (and between themselves) within the uncertainties, the error bars are relatively sizeable, specially in the tiles scheme and for large masses. It should be noted that the error bars from \cite{CUE21} (where the tiles scheme was used) were substantially smaller due to fixing the HOD distribution for the final conclusions. No significant differences are found between both approaches in this non-tomographic case (aside from a slight tendency toward larger values of the HMF and the smaller overall uncertainties for the minitiles scheme), which justifies the use of the tiles scheme in the preliminary work by \cite{CUE21}, which meant to provide a proof-of-concept method to observationally constrain the HMF.

\subsection{Tomographic analysis}

\begin{table*}[h]
\caption{Parameter prior distributions and statistical results for the for the 3-parameter tomographic run.}
\centering
\begin{tabular}{c c c c c c c}
\hline
\hline
Parameter & Prior & Median & Mean & $68\%$ CI& $95\%$ CI\\
\hline 
$A$&$\mathcal{U}$[0,1]&$0.41$&$0.40$&$[0.33,0.52]$&$[0.21,0.56]$\\
$a$&$\mathcal{U}$[0,10]&$2.45$&$2.51$&$[1.76,3.24]$&$[1.93,4.33]$\\
$p$&$\mathcal{U}$$[-10,0.5]$&$-1.09$&$-1.05$&$[-1.64,-0.61]$&$[-2.00,0.12]$\\
\hline
$\alpha_1$&$\mathcal{U}$$[0,1.5]$&$0.62$&$0.65$&$[0.00,0.85]$&$[0.00,1.50]$\\
$\text{log}{M}_{min_1}$&$\mathcal{U}$$[9,16]$&$11.03$&$10.83$&$[10.44,11.69]$&$[9.29,11.82]$\\
$\text{log}{M}_{1_1}$&$\mathcal{U}$[9,16]&$13.59$&$13.52$&$[12.77,16.00]$&$[10.88,16.00]$\\
\hline

$\alpha_2$&$\mathcal{U}$$[0,1.5]$&$0.89$&$0.79$&$[0.00,1.05]$&$[0.00,1.50]$\\
$\text{log}{M}_{min_2}$&$\mathcal{U}$$[9,16]$&$12.15$&$11.71$&$[11.54,12.71]$&$[9.00,12.44]$\\
$\text{log}{M}_{1_2}$&$\mathcal{U}$[9,16]&$13.07$&$13.04$&$[12.37,16.00]$&$[10.12,16.00]$\\
\hline

$\alpha_3$&$\mathcal{U}$$[0,1.5]$&$1.34$&$1.23$&$[1.25,1.50]$&$[0.00,1.50]$\\
$\text{log}{M}_{min_3}$&$\mathcal{U}$$[9,16]$&$12.50$&$12.42$&$[12.17,12.94]$&$[11.46,13.22]$\\
$\text{log}{M}_{1_3}$&$\mathcal{U}$[9,16]&$12.83$&$12.91$&$[12.19,13.40]$&$[11.84,16.00]$\\
\hline

$\alpha_4$&$\mathcal{U}$$[0,1.5]$&$0.71$&$0.75$&$[0.00,1.50]$&$[0.00,1.50]$\\
$\text{log}{M}_{min_4}$&$\mathcal{U}$$[9,16]$&$13.66$&$13.49$&$[13.45,13.89]$&$[11.64,14.19]$\\
$\text{log}{M}_{1_4}$&$\mathcal{U}$[9,16]&$14.73$&$14.60$&$[14.20,16.00]$&$[12.50,16.00]$\\
\hline
\hline
\end{tabular}
\label{table_tomo_allbins_minitiles}
\tablefoot{The column information is the same as the previous tables.}
\end{table*}

Having explored the non-tomographic case, we now turn to the main objective of this work, which is to study the potential improvements that can arise from a tomographic analysis. To do so, and as explained in section 3.3., we assume that all HMF parameters do not vary across redshift bins, that is, that universality of the HMF holds, while we let the HOD evolve. As stressed in section 3.3, only the minitiles scheme will be considered from now on.

\begin{figure}[h]
    \centering
    \includegraphics[width=0.9\columnwidth]{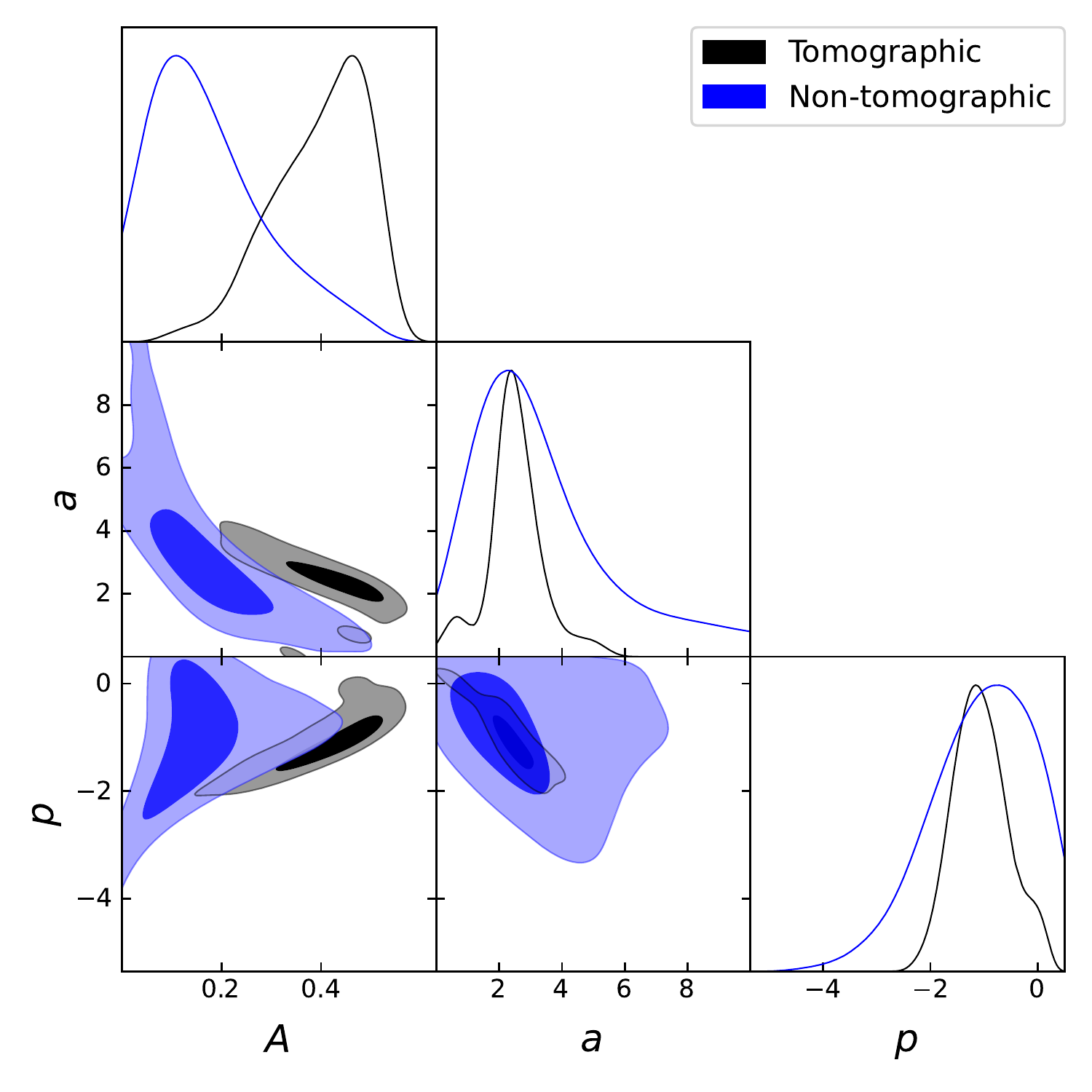}
    \caption{One- and two-dimensional (contour) posterior distributions for the HMF parameters from the tomographic (in black) and non-tomographc (in blue) runs in the minitiles scheme.}
    \label{hmfcorner_tomo_vs_no_tomo_minitiles}
\end{figure}

\begin{figure*}[t]
\centering
\minipage{0.32\textwidth}
  \includegraphics[width=\linewidth]{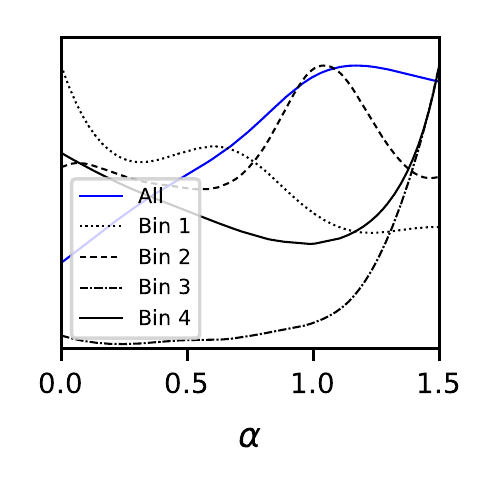}
\endminipage\hfill
\minipage{0.32\textwidth}
  \includegraphics[width=\linewidth]{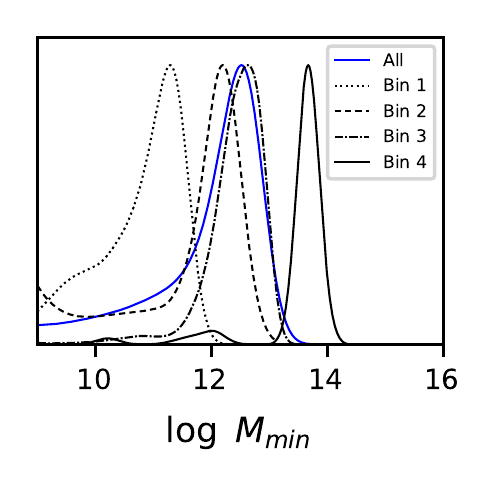}
\endminipage\hfill
\minipage{0.32\textwidth}%
  \includegraphics[width=\linewidth]{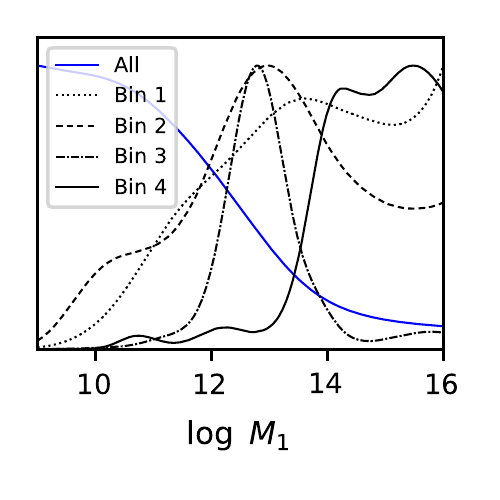}
\endminipage
\caption{One-dimensional posterior distributions for the HOD parameters in the tomographic case under the assumption of universality (in black)
and in the non-tomographic case (in blue).}
\label{1D_tomo_vs_no_tomo_allbins_HOD_minitiles}
\end{figure*}

Table \ref{table_tomo_allbins_minitiles} shows the summarized statistical results from this tomographic run, compared with with the results from the non-tomographic run (in blue). The corresponding part of the full corner plot regarding only the HMF parameters is shown in Fig. \ref{hmfcorner_tomo_vs_no_tomo_minitiles} (in blue) and the marginalized posterior distributions of the HOD are depicted in Fig. \ref{1D_tomo_vs_no_tomo_allbins_HOD_minitiles} (with a different black line for each redshift bin). Figure \ref{corner_tomo_vs_no_tomo} (in black) shows the full corner plot compared with the results from the non-tomographic run (in blue).

The improvement with respect to the non-tomographic case is clear-cut. Indeed, as shown by Fig. \ref{hmfcorner_tomo_vs_no_tomo_minitiles}, the marginalized posterior distributions of $a$ and $p$ are significantly narrowed when compared to the analysis with a single redshift bin. More specifically, $a=2.51^{+0.73}_{-0.75}$ and $p=-1.05^{+0.44}_{-0.59}$, with median values of 2.45 and -1.09, respectively. It should be mentioned that the non-negligible probability region at around $a< 1$ (also visible in the blobs on the $A-a$ plane) is not due to a lack of convergence of the MCMC chains, but to a small number of walkers that remain near those values, as has been confirmed by analyzing their path in parameter space\footnote{Given the condition, addressed in Section 2.1, that the HMF should have a clear physical interpretation, a sufficient number a walkers need to be initialized in finite-probability regions of parameter space, which in our case allowed a small subset of them to stay around those particular values.}. In fact, these values are in the neighborhood of the traditional ST triple.
Concerning the marginalized distribution of the $A$ parameter, it is practically neither narrowed nor widened with respect to the non-tomographic case, but clearly moves toward larger values: $A=0.40^{+0.12}_{-0.07}$. The overall behavior of the HMF parameters can already hint at universality as a very valid assumption for our redshifts, given that the new $a$ and $p$ distributions emphasize the goodness of the fit, but, most importantly, this run clearly shows a substantial improvement regarding uncertainties.

As for the behavior of the HOD parameters with respect to redshift, Fig. \ref{1D_tomo_vs_no_tomo_allbins_HOD_minitiles} shows a clear redshift evolution of the mean minimum halo mass, with larger values of $M_{\text{min}}$ corresponding to earlier times. Particularly, the mean and 68\% credible intervals for $\log{M}_{\text{min}}$ are $10.83^{+0.86}_{-0.39}$, $11.71^{+1.00}_{-0.17}$, $12.42^{+0.52}_{-0.25}$ and $13.49^{+0.40}_{-0.04}$ for bins 1 to 4, respectively. This is the expected behavior \citep{gonzaleznuevo17,BON21}: given that we are working with a flux-limited sample, there is an observational bias toward intrinsically brighter (and thus more massive) objects. With regard to $M_1$, the results for bins 1 and 2 do not show the corresponding evolution, while those from bins 3 and 4 do, with mean values for $\log{M_1}$ of $12.91^{+0.49}_{-0.72}$ and $14.60^{+1.40}_{-0.40}$, respectively. As for the $\alpha$ parameter, it can only be assigned a proper lower bound in bin 3, being $>1.25$ at 68\%, while its overall redshift behavior remains unclear.

\begin{figure}[h]
    \centering
    \includegraphics[width=0.9\columnwidth]{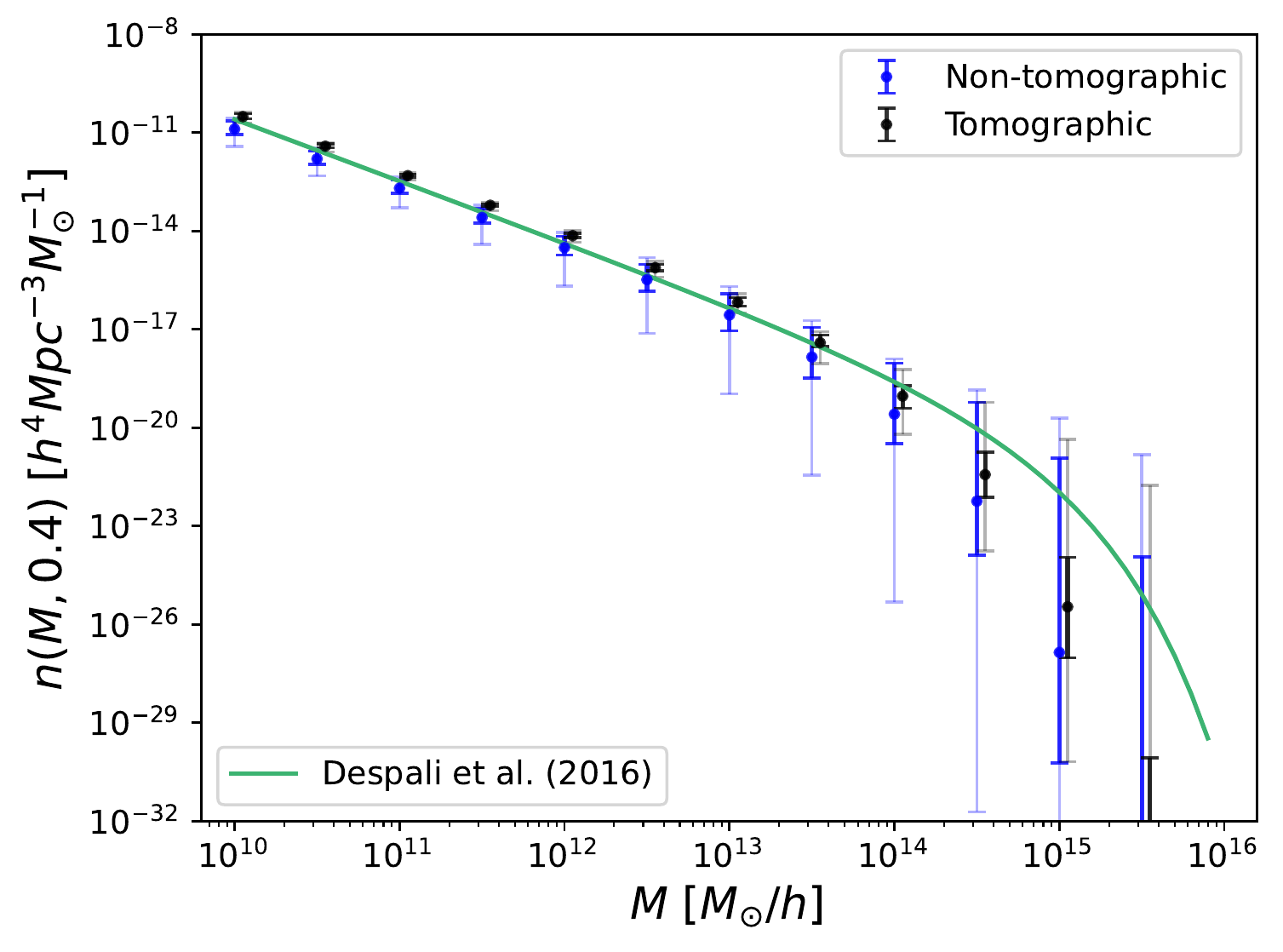}
    \caption{Credible intervals (68\% in bold and 95\% in faint colors) for the $z=0.4$ HMF at different mass values when the full posterior distribution is sampled in both the non-tomographic (blue) and tomographic (black) cases for the minitiles scheme. The ST best fit by \cite{despali16} is drawn with a green line.}
    \label{tabulated_hmf_tomo_vs_no_tomo}
\end{figure}

Fig. \ref{tabulated_hmf_tomo_vs_no_tomo} shows the corresponding median values and credible intervals for the $z=0.4$ HMF and brings to light the remarkable improvements in the actual observational constraints of the HMF. As expected, the uncertainties are substantially smaller with respect to the non-tomographic case, even for 95\% credibility. The error bars for small masses (below $\sim 10^{12} M_{\odot}/h$) show that our results are not compatible with the best-fit ST values found by \cite{despali16} at 95\% credibility. In fact, under the assumption that the HMF is well fit by the ST model, our results predict a larger number density of low-mass halos and (although only at 1$\sigma$) a smaller number of massive ones, that is, a steeper fall with a cutoff at $\sim 10^{14} M_{\odot}/h$.

\begin{figure}[h]
    \centering
    \includegraphics[width=0.9\columnwidth]{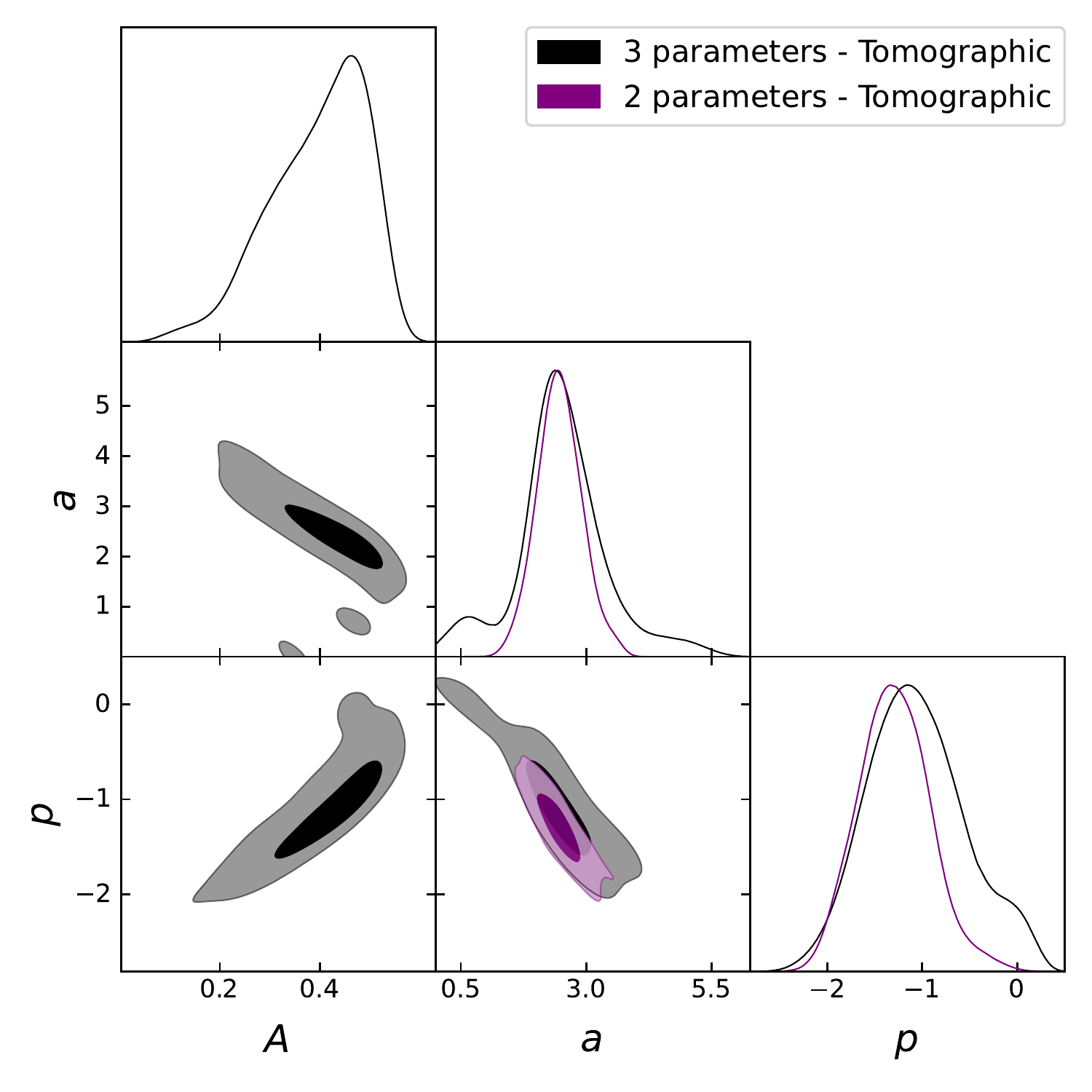}
    \caption{One- and two-dimensional (contour) posterior distributions for the HMF parameters in the tomographic setup for the three-parameter (in black) and two-parameter (in purple) fits.}
    \label{hmfcorner_tomo_3param_vs_2param}
\end{figure}

At this point, it is of interest to analyze the potential of a two-parameter ST fit in the tomographic frame, that is, an ST fit where the $A$ parameter is fixed according to the normalization condition mentioned in Section 2.1. As discussed in \citet{CUE21}, this case is not suitable in a non-tomographic setting due to important degeneracies, a conclusion that has been further confirmed for this work with several tests that yielded nonrestrictive constraints on the $a$ parameter. However, it might be the case that this problem is solved via tomography. Table \ref{table_tomo_allbins_minitiles} shows the summarized statistical results from this tomographic run. The corresponding part of the full corner plot regarding only the HMF parameters is shown in Fig. \ref{hmfcorner_tomo_3param_vs_2param}. Fig. \ref{corner_tomo_3param_vs_2param} depicts the full corner plot from this run, comparing the results with those from the three-parameter case.

\begin{table*}[h]
\caption{Parameter prior distributions and statistical results for the for the two-parameter tomographic run.}
\centering
\begin{tabular}{c c c c c c c}
\hline
\hline
Parameter & Prior & Median & Mean & $68\%$ CI& $95\%$ CI\\
\hline 
$a$&$\mathcal{U}$[0,10]&$2.45$&$2.47$&$[2.01,2.90]$&$[1.59,3.44]$\\
$p$&$\mathcal{U}$$[-10,0.5]$&$-1.30$&$-1.29$&$[-1.66,-0.93]$&$[-2.05,-0.56]$\\
\hline
$\alpha_1$&$\mathcal{U}$$[0,1.5]$&$0.57$&$0.61$&$[0.00,0.79]$&$[0.00,1.50]$\\
$\text{log}{M}_{min_1}$&$\mathcal{U}$$[9,16]$&$11.02$&$10.77$&$[10.37,11.72]$&$[9.00,11.61]$\\
$\text{log}{M}_{1_1}$&$\mathcal{U}$[9,16]&$13.39$&$13.29$&$[12.44,16.00]$&$[10.56,16.00]$\\
\hline

$\alpha_2$&$\mathcal{U}$$[0,1.5]$&$0.94$&$0.87$&$[0.66,1.39]$&$[0.00,1.50]$\\
$\text{log}{M}_{min_2}$&$\mathcal{U}$$[9,16]$&$12.09$&$11.64$&$[11.42,12.67]$&$[9.00,12.41]$\\
$\text{log}{M}_{1_2}$&$\mathcal{U}$[9,16]&$13.14$&$13.17$&$[11.68,15.19]$&$[10.48,16.00]$\\
\hline

$\alpha_3$&$\mathcal{U}$$[0,1.5]$&$1.39$&$1.33$&$[1.35,1.50]$&$[0.94,1.50]$\\
$\text{log}{M}_{min_3}$&$\mathcal{U}$$[9,16]$&$10.50$&$10.72$&$[9.00,11.31]$&$[9.00,12.81]$\\
$\text{log}{M}_{1_3}$&$\mathcal{U}$[9,16]&$11.03$&$11.21$&$[9.70,12.10]$&$[9.05,13.35]$\\
\hline

$\alpha_4$&$\mathcal{U}$$[0,1.5]$&$0.72$&$0.74$&$[0.00,1.50]$&$[0.00,1.50]$\\
$\text{log}{M}_{min_4}$&$\mathcal{U}$$[9,16]$&$13.64$&$13.63$&$[13.56,13.73]$&$[13.46,13.80]$\\
$\text{log}{M}_{1_4}$&$\mathcal{U}$[9,16]&$14.64$&$14.71$&$[14.21,16.00]$&$[13.69,16.00]$\\
\hline
\hline
\end{tabular}
\label{table_tomo_caseB}
\tablefoot{The column information is the same as the previous tables.}
\end{table*}

The most remarkable result is the fact that constraints on the HMF parameters are even tighter with respect to the three-parameter case and, surprisingly, that the difficulties present in the two-parameter non-tomographic case are surmounted by the adoption of the tomographic frame. In particular we estimate the following values for the HMF parameters: $a=2.47^{+0.43}_{0.46}$ and $p=-1.29^{+0.36}_{-0.37}$, with median values of 2.45 and -1.30, respectively. The traditional ST values are rejected at more than $3\sigma$, as will be later discussed. 

As for the HOD parameters, Fig. \ref{corner_tomo_3param_vs_2param} depicts a qualitatively similar situation to the three-parameter case except for a large discrepancy in the third bin. Indeed, the $M_{\text{min}}$ parameter shows the expected redshift evolution in bins 1, 2 and 4, with mean values for $\log{M_{\text{min}}}$ of $10.75^{+1.08}_{-0.60}$, $11.38^{+1.27}_{-0.59}$ and $13.12^{+0.97}_{-0.06}$, respectively. However, the one-dimensional distribution for bin 3 displays an unusually large probability region for low masses, which breaks this tendency. We believe this effect is caused by the seemingly preferred low values of $M_1$ in this bin, which in turn forces $M_{\text{min}}$ to be smaller. Regardless of this feature, it should be mentioned that the marginalized posterior distributions of both $a$ and $p$ converge much more quickly than those of the HOD parameters, which proves the robustness of these results with respect to the HOD. Moreover, the main goal of this work is to provide observational constraints on the HMF, not the precise determination of the HOD.

\begin{figure}[h]
    \centering
    \includegraphics[width=0.9\columnwidth]{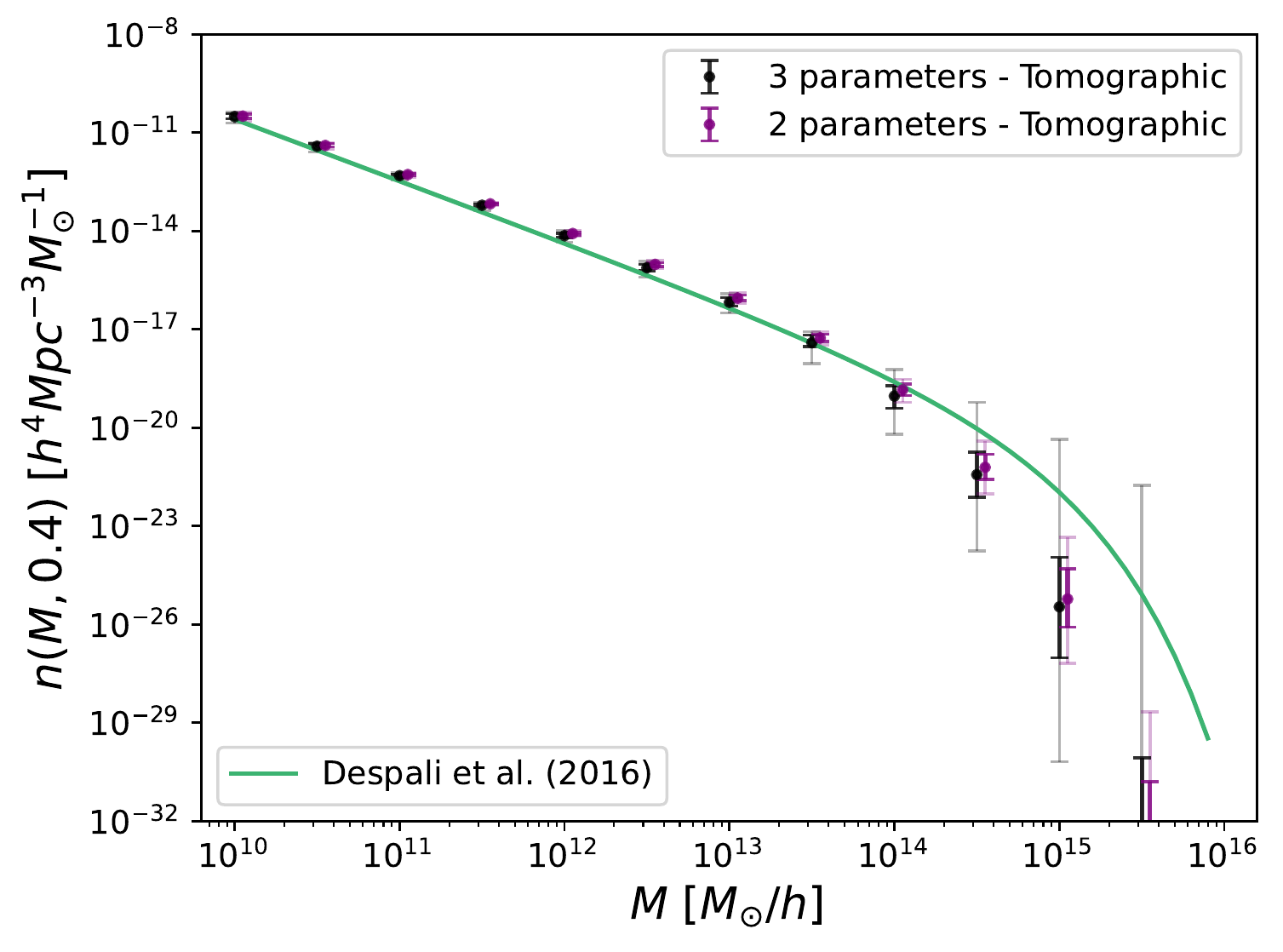}
    \caption{Credible intervals (68\% in bold and 95\% in faint colors) for the $z=0.4$ HMF at different mass values when the full posterior distribution is sampled in the tomographic setup for the three-parameter (in black) and the two-parameter (in purple) fits. The ST best fit by \cite{despali16} is drawn with a green line.}
    \label{tabulated_hmf_tomo_3param_vs_2param}
\end{figure}

Fig. \ref{tabulated_hmf_tomo_3param_vs_2param} shows the median values and credible intervals for the $z=0.4$ HMF obtained for this run (in purple) and compared to those from the three-parameter fit (in black) and the ST best fit found by \cite{despali16} (in green). As commented before and stressed by this figure, the disagreement between our results for the two-parameter case and the best-fit ST values is stronger than in the previous scenario, even at large masses and it can safely be said that we predict a higher number density of halos for masses below $\sim 10^{13} M_{\odot}/h$ and lower number density for masses above $\sim 10^{14}M_{\odot}/h$, and thus a steeper fall in mass.

\subsection{A step toward testing universality}

\begin{table*}[t]
\caption{Parameter prior distributions and statistical results for the for the three-parameter non-tomographic runs.}
\centering
\begin{tabular}{c c c c c c c}
\hline
\hline
Parameter & Prior & Median & Mean & $68\%$ CI& $95\%$ CI\\
\hline 
$A_1$&$\mathcal{U}$[0,1]&$0.28$&$0.29$&$[0.15,0.41]$&$[0.07,0.52]$\\
$a_1$&$\mathcal{U}$[0,10]&$6.11$&$5.91$&$[4.61,10.00]$&$[0.00,10.00]$\\
$p_1$&$\mathcal{U}$$[-10,0.5]$&$-0.57$&$-0.70$&$[-1.13,0.23]$&$[-2.05,0.49]$\\
$\alpha_1$&$\mathcal{U}$$[0,1.5]$&$0.88$&$0.85$&$[0.63,1.50]$&$[0.19,1.50]$\\
$\text{log}{M}_{min_1}$&$\mathcal{U}$$[9,16]$&$11.22$&$10.97$&$[10.53,11.99]$&$[9.07,12.09]$\\
$\text{log}{M}_{1_1}$&$\mathcal{U}$[9,16]&$13.11$&$13.09$&$[12.12,16.00]$&$[10.23,16.00]$\\
\hline

$A_2$&$\mathcal{U}$[0,1]&$0.25$&$0.26$&$[0.10,0.38]$&$[0.02,0.50]$\\
$a_2$&$\mathcal{U}$[0,10]&$2.89$&$3.37$&$[0.48,4.37]$&$[0.00,8.07]$\\
$p_2$&$\mathcal{U}$$[-10,0.5]$&$-0.80$&$-0,93$&$[-1.45,0.18]$&$[-2.54,0.50]$\\
$\alpha_2$&$\mathcal{U}$$[0,1.5]$&$1.04$&$0.95$&$[0.80,1.50]$&$[0.23,1.50]$\\
$\text{log}{M}_{min_2}$&$\mathcal{U}$$[9,16]$&$12.10$&$11.60$&$[11.13,12.90]$&$[9.00,12.63]$\\
$\text{log}{M}_{1_2}$&$\mathcal{U}$[9,16]&$13.02$&$12.96$&$[11.35,15.00]$&$[10.10,16.00]$\\
\hline

$A_3$&$\mathcal{U}$[0,1]&$0.45$&$0.40$&$[0.36,0.55]$&$[0.10,0.58]$\\
$a_3$&$\mathcal{U}$[0,10]&$1.99$&$2.14$&$[1.31,2.79]$&$[0.00,3.84]$\\
$p_3$&$\mathcal{U}$$[-10,0.5]$&$-0.75$&$-0.89$&$[-1.33,0.07]$&$[-2.45,0.33]$\\
$\alpha_3$&$\mathcal{U}$$[0,1.5]$&$1.41$&$1.35$&$[1.35,1.50]$&$[0.96,1.50]$\\
$\text{log}{M}_{min_3}$&$\mathcal{U}$$[9,16]$&$11.05$&$11.03$&$[9.00,12.00]$&$[9.00,12.84]$\\
$\text{log}{M}_{1_3}$&$\mathcal{U}$[9,16]&$11.45$&$11.64$&$[9.88,12.72]$&$[9.08,14.99]$\\
\hline

$A_4$&$\mathcal{U}$[0,1]&$0.24$&$0.26$&$[0.10,0.36]$&$[0.03,0.50]$\\
$a_4$&$\mathcal{U}$[0,10]&$4.86$&$5.17$&$[2.45,7.30]$&$[0.00,10.00]$\\
$p_4$&$\mathcal{U}$$[-10,0.5]$&$-0.77$&$-0.91$&$[-1.40,0.26]$&$[-2.55,0.50]$\\
$\alpha_4$&$\mathcal{U}$$[0,1.5]$&$0.79$&$0.79$&$[0.53,1.50]$&$[0.00,1.50]$\\
$\text{log}{M}_{min_4}$&$\mathcal{U}$$[9,16]$&$13.69$&$13.58$&$[13.49,13.89]$&$[12.75,14.18]$\\
$\text{log}{M}_{1_4}$&$\mathcal{U}$[9,16]&$14.64$&$14.63$&$[13.95,15.54]$&$[13.69,16.00]$\\
\hline
\hline
\end{tabular}
\label{table_tomo_no_universal}
\tablefoot{The column information is the same as the previous tables.}
\end{table*}

Since we have measured the cross-correlation function for each foreground bin, we can now explore the possibility of allowing the HMF parameters to vary with redshift and perform four different non-tomographic MCMC runs, one for each bin. It should be noticed that it is well established that universality in the HMF strongly depends on the mass definition of halos \citep{courtin11,despali16,ONDARO21}, which means that our current results in this area should be regarded as very preliminary, since a serious study concerning differences with regard to mass definition is beyond the scope of this work. 

Taking into account that the two-parameter fit does not perform well in a non-tomographic setting, we carried out these runs for the three-parameter case only.Table \ref{table_tomo_no_universal} shows the summarized statistical results from these runs. For visual purposes, the marginalized posterior distributions of the HOD and the HMF parameters in each redshift bin are shown in Fig. \ref{1D_tomo_HMF_and_HOD_no_universality}, while Fig. \ref{tomo_caseB_cornerplot} depicts the full corner plot.

If we focus first on the HOD parameters, their marginalized posterior distributions surprisingly show the same distinctive feature as the two-parameter fit in the tomographic case. Indeed, both the $M_{\text{min}}$ and $M_1$ parameters take unusually low values with very high probability in bin 3. The reason why this happens in this three-parameter case is not clear as of now (since the $A_3$ parameter is given freedom as opposed to the two-parameter scenario), but we stress that the main goal of this work is to constrain the HMF and not give tight bounds on the HOD; nevertheless, this issue will be studied in the future. As for the rest of bins, the behavior is completely similar to that of the tomographic case, with quantitative differences mainly of the form of a larger dispersion for bins 1 and 2.  
The situation is more interesting for the HMF parameters. With respect to $p$, Fig. \ref{1D_tomo_HMF_and_HOD_no_universality} shows a perfect non-evolving probability distribution, with barely any difference among bins. As for $A$, the situation differs due again to bin 3. While the posterior distributions for the rest bins 1, 2, and 4 remain practically unchanged, that of bin 3 is displaced to notoriously larger values with a low-probability tail at the mode of the other distributions. This is again indicative of a characteristic behavior of bin 3, which however translates into a smaller spread in the distribution (and consequently, in the HMF). This is patently obvious in the marginalized distributions of the $a$ parameter. Bin 1 displays an ever-increasing probability distribution, a behavior utterly opposed to that of the rest of bins, probably due to the poor cross-correlation data at this redshift. As for the $a$ parameter in bin 3, it displays a relatively narrow probability distribution when compared to the rest, which keep the same tendency but with larger uncertainties.

\setcounter{figure}{12}
\begin{figure}[h]
    \centering
    \includegraphics[width=0.9\columnwidth]{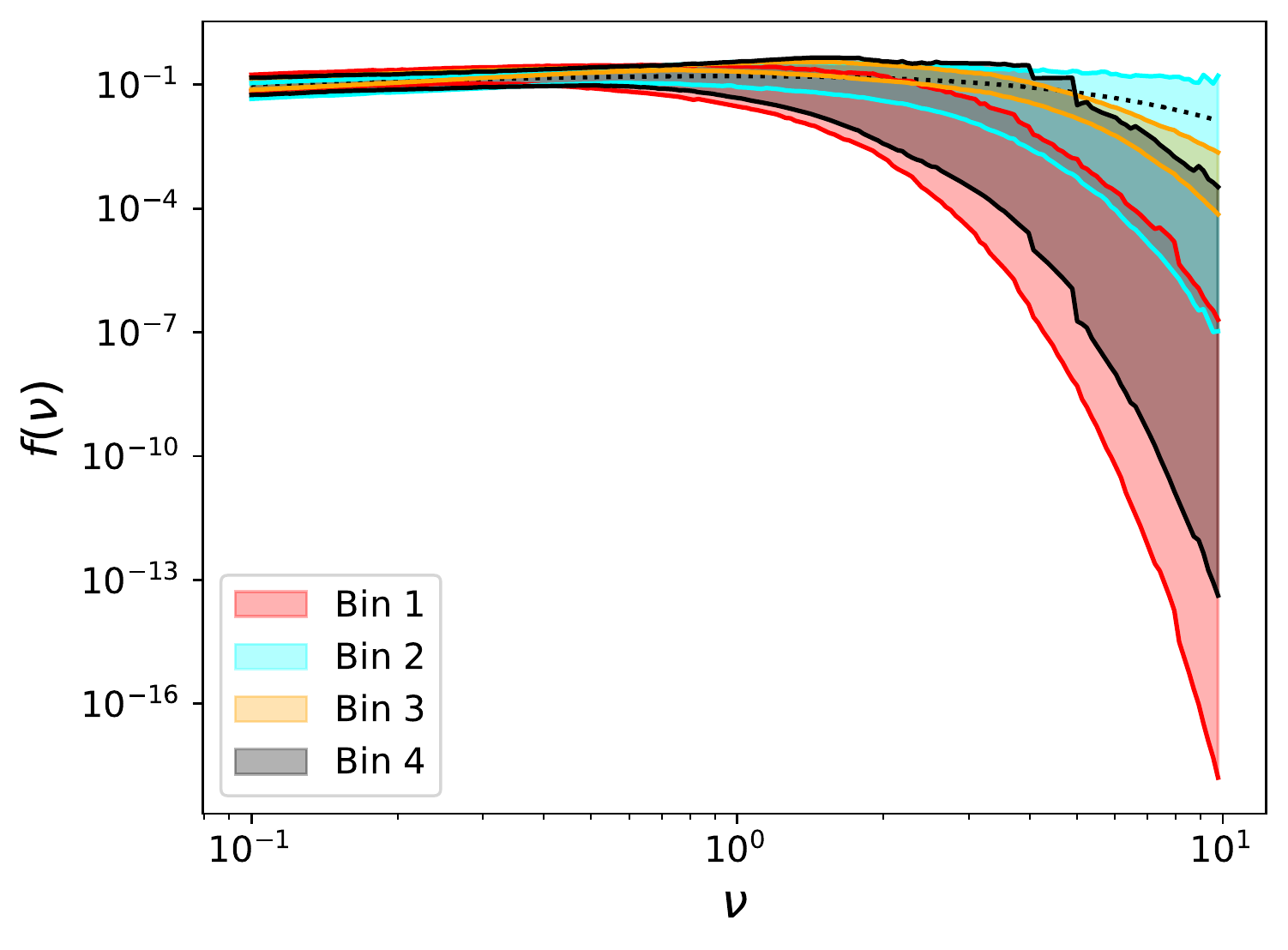}
    \caption{Posterior-sampled $f(\nu,z)$ function at different redshift bins. The 68$\%$ probability contours are depicted in red, cyan, orange, and black for bins 1 to 4, respectively. The ST best fit by \cite{despali16}  is plotted with a black dotted line.}
    \label{universality}
\end{figure}

\setcounter{figure}{13}
\begin{figure*}[h]
\centering
\minipage{0.32\textwidth}
  \includegraphics[width=\linewidth]{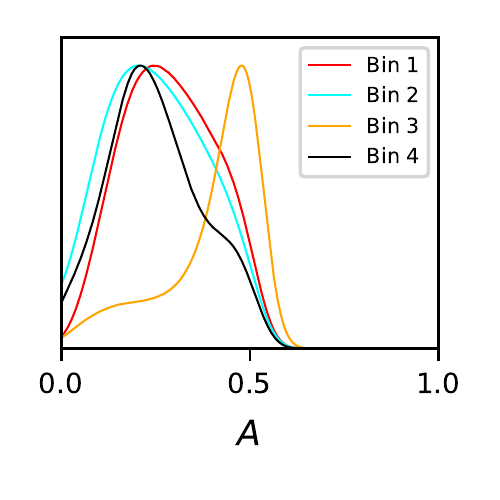}
\endminipage\hfill
\minipage{0.32\textwidth}
  \includegraphics[width=\linewidth]{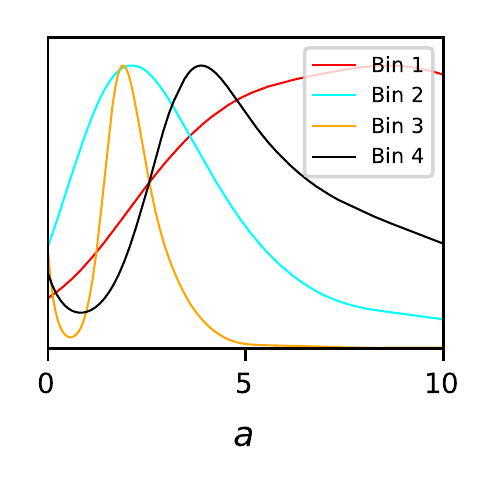}
\endminipage\hfill
\minipage{0.32\textwidth}%
  \includegraphics[width=\linewidth]{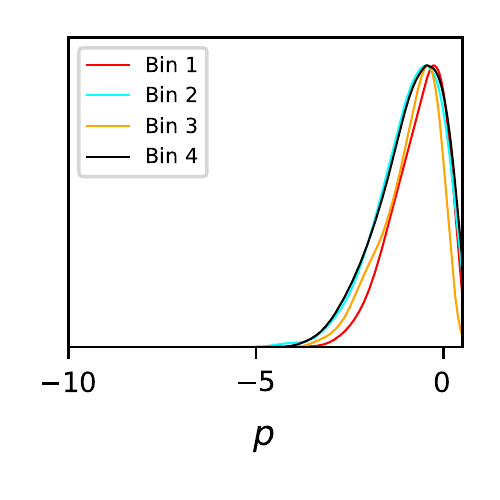}
\endminipage\\
\minipage{0.32\textwidth}
  \includegraphics[width=\linewidth]{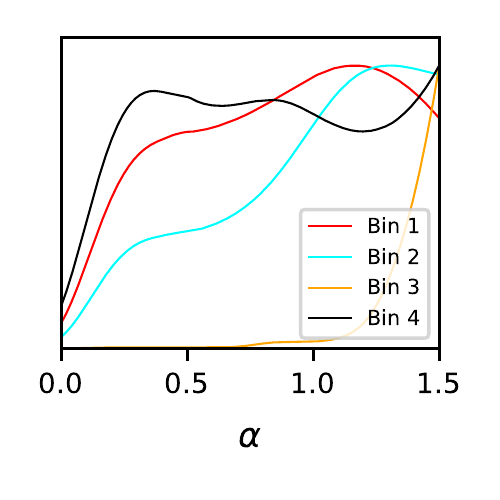}
\endminipage\hfill
\minipage{0.32\textwidth}
  \includegraphics[width=\linewidth]{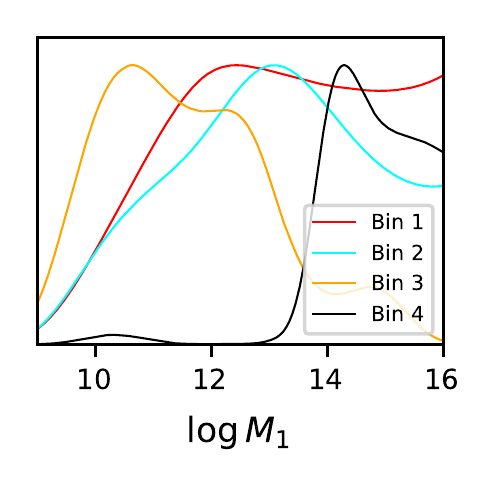}
\endminipage\hfill
\minipage{0.32\textwidth}%
  \includegraphics[width=\linewidth]{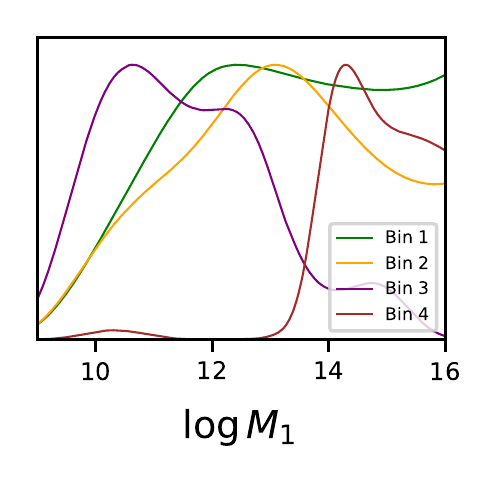}
\endminipage

\caption{One-dimensional posterior distributions for the HMF (top panel) and HOD (bottom panel) parameters in the last four nontomographic runs. The results from bins 1, 2, 3, and 4 are shown with a red, cyan, orange and black line, respectively.}
\label{1D_tomo_HMF_and_HOD_no_universality}
\end{figure*}

Given the very large overlap in the distributions, universality seems to hold at first glance. However, to test it in a quantitative manner, we sampled the function $f(\nu,z)$ from \eqref{sthmf} using the posterior distributions from all redshift bins, whose corresponding $68\%$ probability bands are shown in Fig. \ref{universality}, along with the ST best fit found by \cite{despali16} (black dotted line). The most obvious aspect to be noted is that the contour from bin 3 is extremely narrow with respect to the rest, which are patently wider due to the long high-probability tails of the marginalized distribution of the $a$ parameter, which sources most of the HMF uncertainty. This causes that these three contours overlap at every value of $\nu$, but those of bin 3 and bin 1 do not at 68\% probability. However, as stressed at the start of this section and given the large uncertainties, this should be more thoroughly studied in the future by varying the halo mass definition and performing parallel tomographic analysis in different sets of bins.


\section{Summary and conclusions}
This paper has addressed the issue of observationally restricting the dark matter halo mass function by measuring the cross-correlation function between samples of GAMA II (with spectroscopic redshifts between $0.1<z<0.8$) and H-ATLAS galaxies (with photometric redshifts between $1.2<z<4.0$) in a tomographic analysis. Due to their physical properties, submillimeter galaxies have been chosen as background sources so as to optimize the measurement of the signal, which is a clear manifestation of the phenomenon of magnification bias. By means of a halo model interpretation of the cross-correlation function and the ST parametrization of the HMF, a Bayesian analysis through an MCMC algorithm has been carried out to obtain posterior probability distributions for both the HMF and HOD parameters under a fixed \textit{Planck} cosmology and credible intervals for the number density of the dark matter halos associated with the lenses at given mass values. Two different tiling schemes, namely the so-called tiles and minitiles schemes, have been initially used for the sake of coherence, although the statistically rigorous results come from using the latter given the larger sample size.

We began the analysis with a non-tomographic study that considered the overall lens sample so as to assess the information that can be extracted without taking into account the possible redshift variation of the HOD parameters. The results from both tiling schemes are qualitatively similar, with slightly smaller overall uncertainties for the minitile case. The observational constraints on the HMF agree very well with the traditional result from numerical simulations, but the error bars are large, specially for halo masses above $\sim 10^{14} M_{\odot}/h$. 

In order to explore the potential improvements provided by dividing the foreground sample, we performed a tomographic analysis under the assumption of universality of the HMF and the three-parameter ST fit in the minitiles scheme. The main result of this paper is the remarkable improvement in the observational constraints of the HMF when adopting a tomographic setup, which clearly confirms the forecast by \cite{CUE21}. Under the universal three-parameter ST fit, and with respect to the traditional values from numerical simulations, we predict a higher number density of dark matter halos at masses below $\sim 10^{12} M_{\odot}/h$ (at $95\%$ credibility) and, although at only $1\sigma$, a tendency toward a lower number density at masses larger than $10^{14} M_{\odot}/h$. The ST values found by \cite{despali16} are thus not compatible at 95\% credibility.

Although a two-parameter ST fit did not perform well in a non-tomographic setup \citep[as shown in ][]{CUE21}, we decided to test it in the tomographic case. The results are even more remarkable and further restrict the HMF constraints from the previous case, predicting once more a higher number density of halos for low masses (in this case below $\sim 10^{13} M_{\odot}/h$ and a lower number density above $\sim 10^{14} M_{\odot}/h$, all this at more than $3\sigma$ credibility, thus disagreeing again the traditional ST triples from numerical simulations, this time more strongly. This is worth noticing and a comment should be made. On the one hand, the inconsistency of traditional ST parameter values with the data may challenge
the interpretation of the ST formula in terms of the excursion set theory with an ellipsoidal barrier,
since the original values are strictly related to the barrier shape and to the way the mass function is computed in such a theoretical framework. On the other hand, if the ST formula is taken at face value as a simple fit to results from $N-$body simulations, one might not consider this to be too relevant, especially given the compatibility at 3$\sigma$ of the three-parameter case.

Lastly, and so as to use the cross-correlation measurements in each bin to test the assumption of universality in a very preliminary test, four independent MCMC runs were carried out. While the results for the HOD parameters are similar to those from the tomographic case (except for some characteristic feature in the third redsfhit bin), there appear to be discrepancies between the lenses at redshifts $0.3-0.5$ and the rest, although given the large uncertainties, a more in-depth study regarding halo mass definition should be performed in the future to provide a more thorough analysis and robust conclusions.

Following this line of thought, an enlargement of both the foreground and background samples is needed to improve statistics \citep[as discussed in][]{CUE21} due to the nature of a tomographic analysis, where the number of sources used to compute the cross-correlation is decreased with respect to the non-tomographic case. Another interesting approach would involve parametrizing the HMF with a different model to study variations in our observational constraints. While \cite{CUE21} already performed such an analysis in the non-tomographic case comparing the ST and Tinker \citep{tinker08} fits and concluding that statistical problems arose regarding prior ranges for the latter, it could very well be that these issues would be solved through tomography. Another possibility could of course be the use of other HMF models
derived from alternatives to the standard excursion set formalism, like the stochastic theory  by \citet{LAP20,LAP21}.

Lastly, it should be noted that the halo model used in this work has its shortcomings and appears not to show the accuracy required for very precise measurements regarding cosmology, mainly due to the transition between the 1-halo and 2-halo regimes and/or baryonic physics \citep{MEAD15,Mead20,PHIL20,ACUTO21}. Future work will also address the possible improvements that could arise when the model is refined.

\begin{acknowledgements}

MMC, JGN, LB, DC, JMC acknowledge the PGC 2018 project PGC2018-101948-B-I00 (MICINN/FEDER).
MMC acknowledges PAPI-20-PF-23 and PAPI-21-PF-04 (Universidad de Oviedo).\\
AL is supported by the EU H2020-MSCA-ITN-2019 Project 860744 “BiD4BESt: Big Data applications for black hole Evolution STudies.” and by
the PRIN MUR 2017 prot. 20173ML3WW “Opening the ALMA window on the cosmic evolution of gas, stars and supermassive black holes”.\\
We deeply acknowledge the CINECA award under the ISCRA initiative, for the availability of high performance computing resources and support. In particular the COSMOGAL projects “SIS20\_lapi”, “SIS21\_lapi” in the framework “Convenzione triennale SISSA-CINECA”.\\

The \textit{Herschel}-ATLAS is a project with \textit{Herschel}, which is an ESA space observatory with science instruments provided by European-led Principal Investigator consortia and with important participation from NASA. The H-ATLAS web-site is http://www.h-atlas.org. GAMA is a joint European-Australasian project based around a spectroscopic campaign using the Anglo-Australian Telescope.\\

This research has made use of the python packages \texttt{ipython} \citep{ipython}, \texttt{matplotlib} \citep{matplotlib} and \texttt{Scipy} \citep{scipy}.
\end{acknowledgements}

\newpage
\bibliographystyle{aa} 
\bibliography{./main} 

\appendix

\section{Ingredients of the model}

This section is aimed at giving details about the computation of the theoretical model used in this work. The cross-power spectrum present in the angular cross-correlation function of \eqref{crosscorr} has been calculated via the halo model prescription, that is,
\begin{equation*}
    P_{\text{g-dm}}(k,z)=P_{\text{g-dm}}^{\text{1h}}(k,z)+P_{\text{g-dm}}^{\text{2h}}(k,z),
\end{equation*}
where
\begin{equation*}
    P_{\text{g-dm}}^{\text{1h}}(k,z)=\int_0^{\infty} dM\,M\frac{n(M,z)}{\bar{\rho}(z)}\frac{\langle N_{g}\rangle_M}{\bar{n}_g(z)}|u_{\text{dm}}(k,z|M)||u_{\text{g}}(k,z|M)|^{p-1}
\end{equation*}
and
\begin{align*}
    P_{\text{g-dm}}^{\text{2h}}(k,z)&=P(k,z)\Big[\int_0^{\infty}dM\,M\frac{n(M,z)}{\bar{\rho}(z)}b_1(M,z)u_{\text{dm}}(k,z|M)\Big]\,\cdot\nonumber\\
    &\quad\quad\quad\cdot\Big[\int_0^{\infty}dM\,n(M,z)b_1(M,z)\frac{\langle N_g \rangle_M}{\bar{n}_g(z)}u_g(k,z|M)\Big]
\end{align*}

are the 1-halo and 2-halo term of the cross-power spectrum, respectively. In these expressions, $\bar{\rho}(z)$ is the mean matter density of the Universe at redshift $z$, $n(M,z)$ is the HMF, computed through \eqref{HMFnu} under the ST parametrization, $b_1(M,z)$ is the determinsitic bias, computed through the peak background split as in \cite{st1999}. The mean number of galaxies in a halo of mass $M$, $\langle N_g\rangle_M$ is parametrized as described in \eqref{Ngalaxies} and the mean number density of galaxies at redshit $z$ is thus given by
\begin{equation*}
    \bar{n}_g(z)=\int_0^{M}\,dM\,n(M,z)\,\langle N_g\rangle_M.
\end{equation*}

The linear matter power spectrum $P(k,z)$ is evolved to redshift $z$ through the linear growth factor approximation of \cite{carroll92} and the corresponding transfer function is computed via Eisenstein and Hu's fitting formula \cite{eisenstein98}, which takes baryonic effects into account. We have chosen an analytic approach as opposed to a numerical computation using a Boltzmann solver mainly due to computation time, since no significant differences were found between the two.

The normalized Fourier transform of the dark matter distribution within an NFW halo of mass M is given by \citep{cooray02}
\begin{align*}
    u_{\text{dm}}(k,z|M)&=\frac{4\pi\rho_s r_s^3(M,z)}{M}\Big[\sin{kr_s}\big[\text{Si}([1+c]kr_s)-\text{Si}(kr_s)\big]-\\
    &-\frac{\sin{ckr_s}}{[1+c]kr_s}+\cos{kr_s}\big[\text{Ci}([1+c]kr_s)-\text{Ci}(kr_s)\big]\Big],
\end{align*}
where
\begin{equation}
    r_s(M,z)\equiv \frac{R_{\text{vir}}}{c(M,z)}\label{rs}
\end{equation}
and $\rho_s$ are a scale radius and density that parametrize the profile, concentration parameter of a halo of mass $M$ at redshift $z$, which satisfies
\begin{equation}
    M=4\pi \rho_s r_s^3\Big[\ln{[1+c(M,z)]-\frac{c(M,z)}{1+c(M,z)}}\Big]\label{Mnfw}
\end{equation}
for an NFW profile. The virial radius $R_{\text{vir}}$ was computed via the virial overdensity at redshift $z$, through the fit by \cite{weinberg03}. It should be noted that \cite{st1999} identified halos using a spherical overdensity halo finder, but with an overdensity threshold of 178 times the background density. In practice, for a halo of mass $M$, we adopted the concentration parameter by \cite{bullock01}, computed $r_s$ through \eqref{rs} and, subsequently, computed $\rho_s$ using \eqref{Mnfw}.

Lastly, two additional comments should be made. Firstly, it is a reasonable approximation \citep{shethdiaferio01} to set the Fourier transform of the galaxy distribution, $u_{\text{g}}(k,z)$ equal to that of dark matter. Secondly, the exponent $p$ should be set to 1 for central galaxies and to 2 for satellites.


\onecolumn

\section{Additional plots}

\begin{figure}[h]
    \centering
    \includegraphics[width=0.65\textwidth]{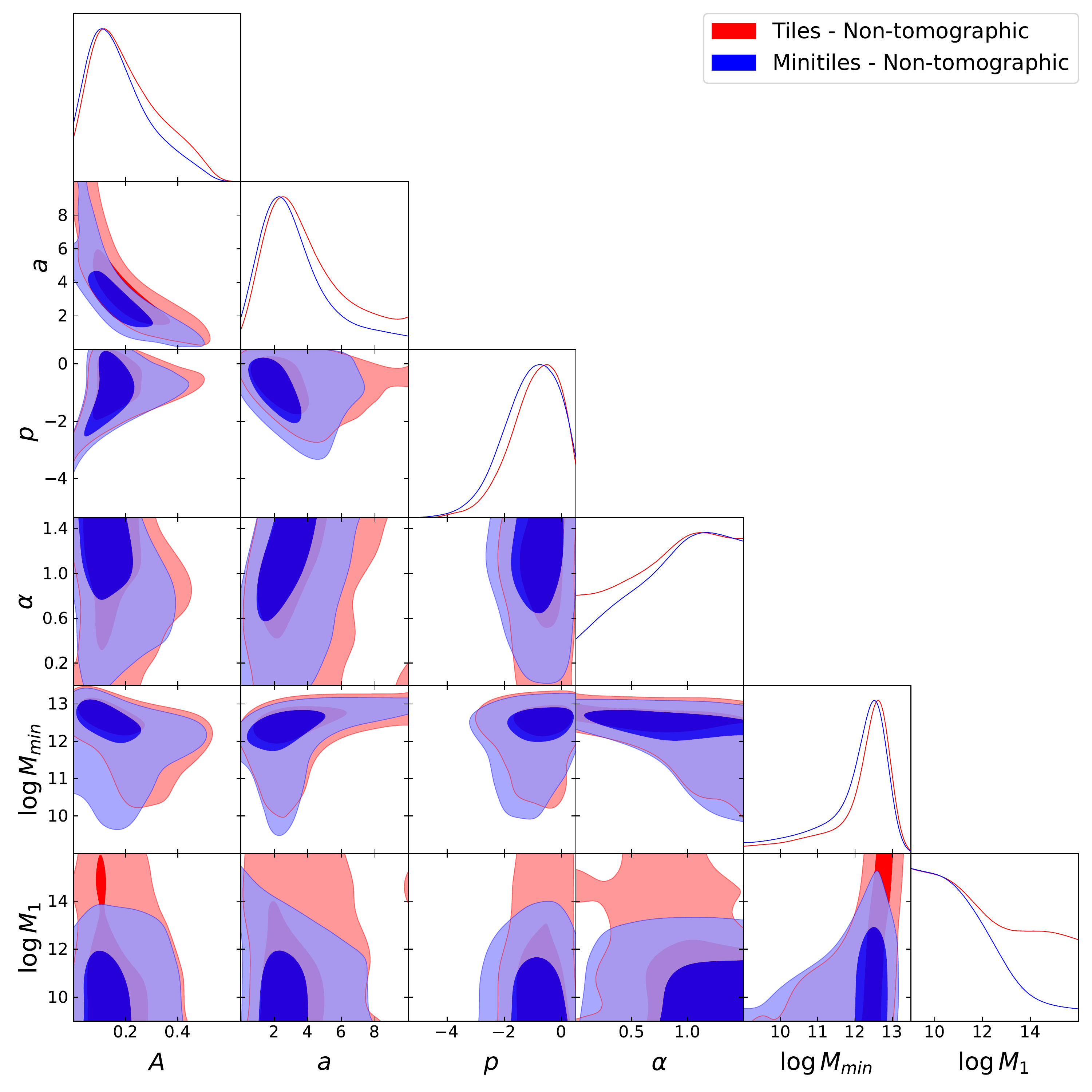}
    \caption{One- and two-dimensional (contour) posterior distributions for the HMF and HOD parameters from the non-tomographic with the tiles (in red) and minitiles (in blue) schemes.}
    \label{corner_nontomo_full_tiles_vs_minitiles}
\end{figure}

\begin{figure*}[h] 
  \begin{minipage}[b]{0.5\linewidth}
    \centering
    \includegraphics[width=0.9\columnwidth]{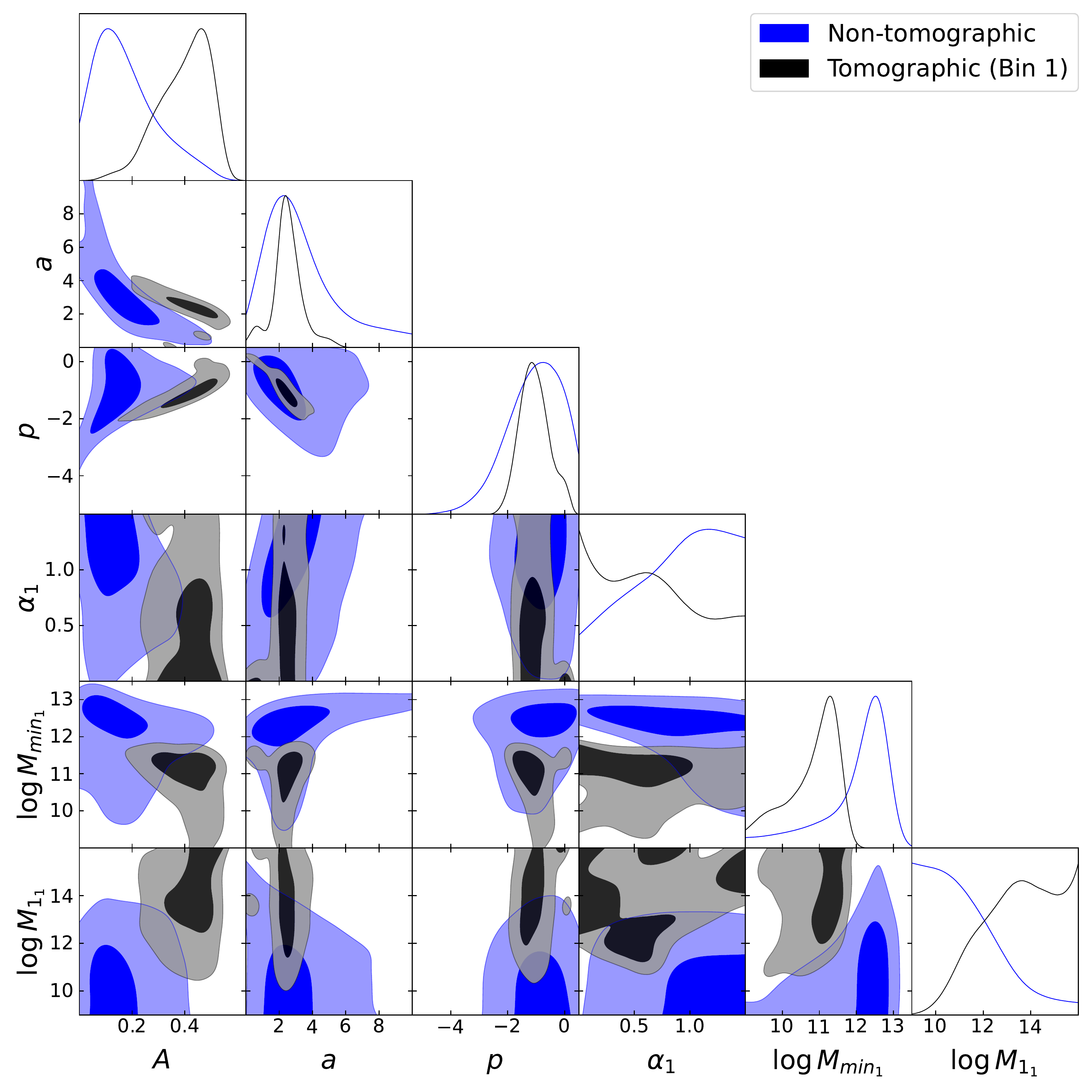} 
  \end{minipage}
  \begin{minipage}[b]{0.5\linewidth}
    \centering
    \includegraphics[width=0.9\columnwidth]{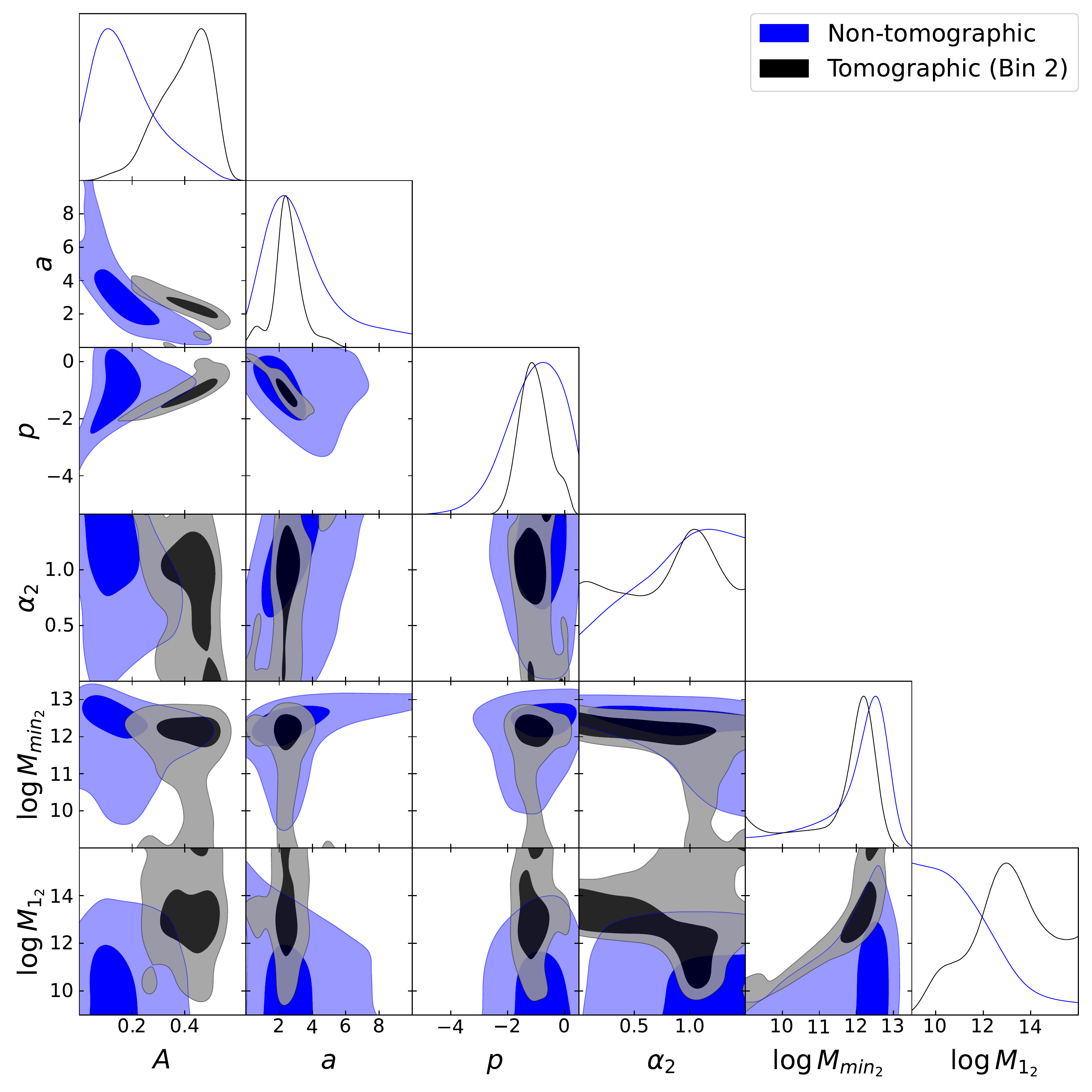} 
  \end{minipage} 
  \begin{minipage}[b]{0.5\linewidth}
    \centering  \includegraphics[width=0.9\columnwidth]{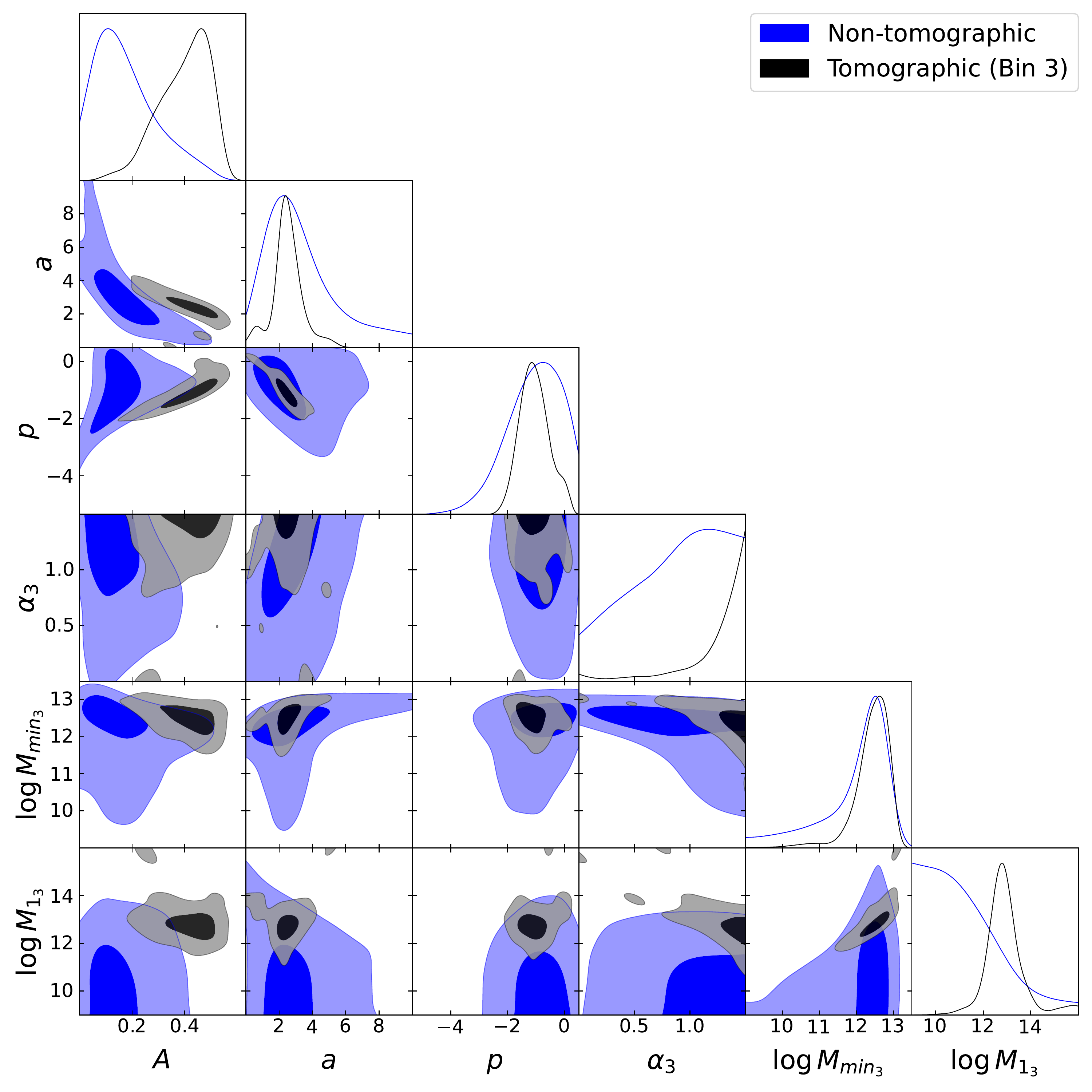}  
  \end{minipage}
  \begin{minipage}[b]{0.5\linewidth}
    \centering
    \includegraphics[width=0.9\columnwidth]{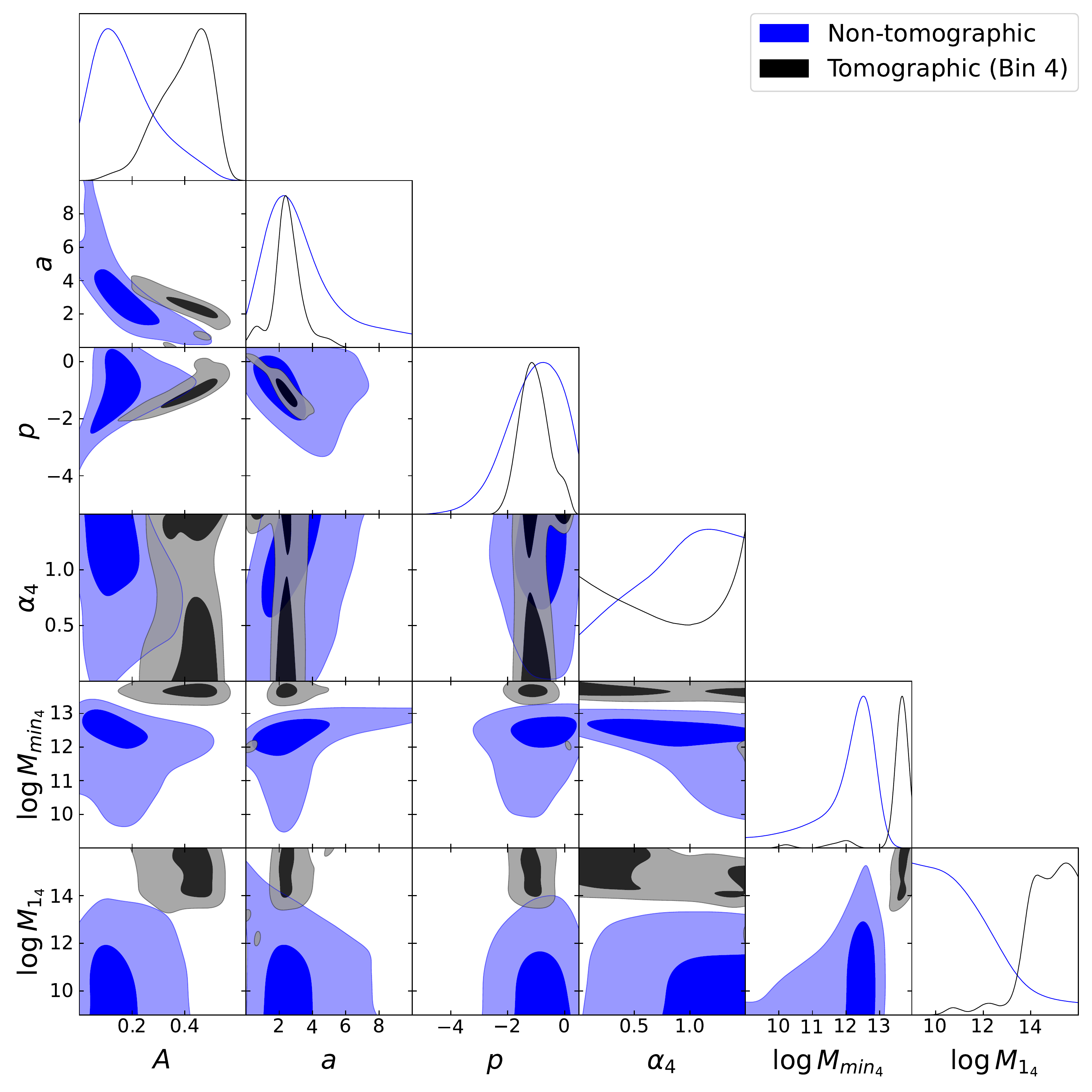}  
  \end{minipage} 
  \caption{One- and two-dimensional (contour) posterior distributions for the HMF and HOD parameters from the tomographic run (in black) and the non-tomographic run (in blue) for the minitiles scheme. The results from bins 1 to 4 are shown, respectively, from left to right and from top to bottom.}
  \label{corner_tomo_vs_no_tomo}
\end{figure*}

\begin{figure*}[h] 
  \begin{minipage}[b]{0.5\linewidth}
    \centering
    \includegraphics[width=0.9\columnwidth]{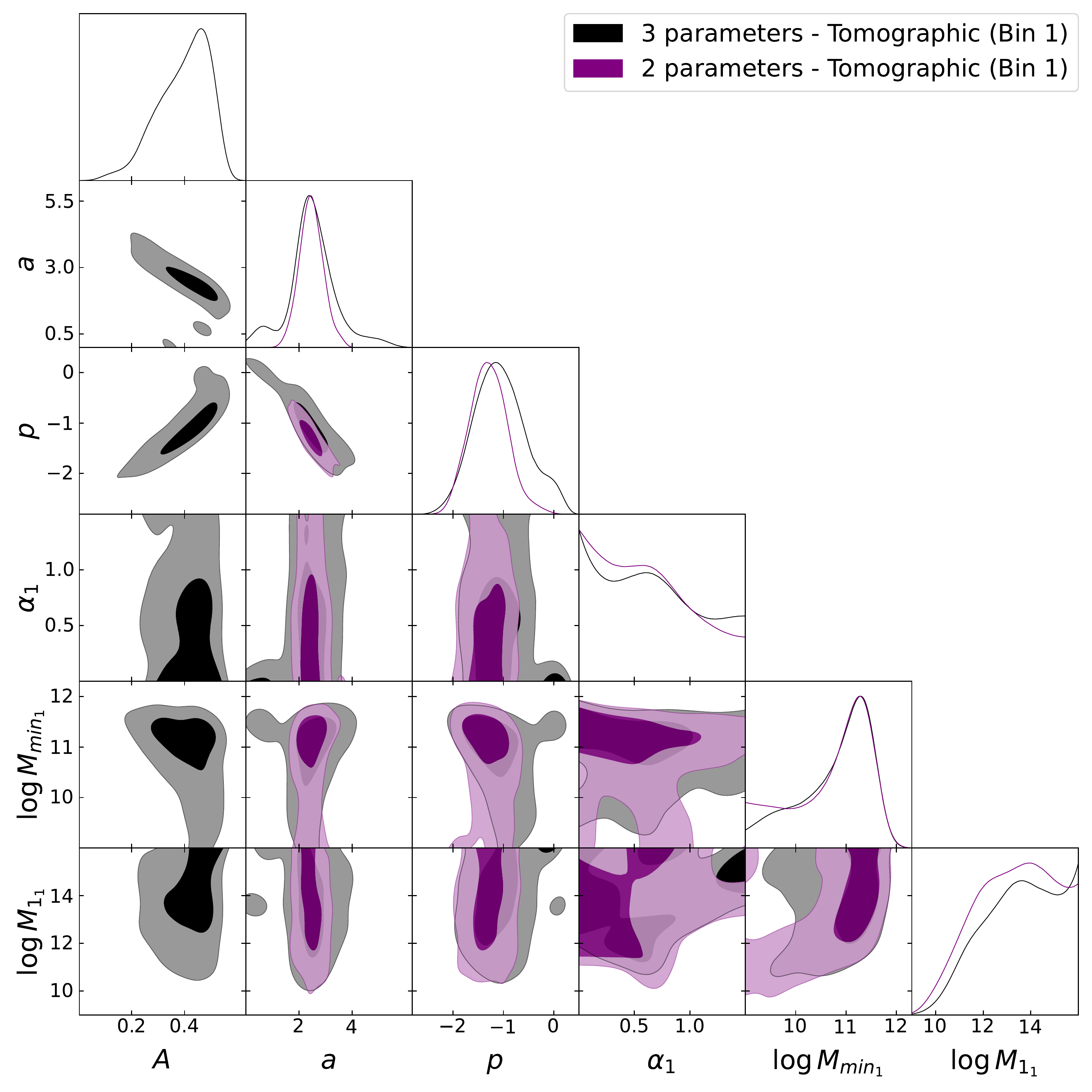} 
  \end{minipage}
  \begin{minipage}[b]{0.5\linewidth}
    \centering
    \includegraphics[width=0.9\columnwidth]{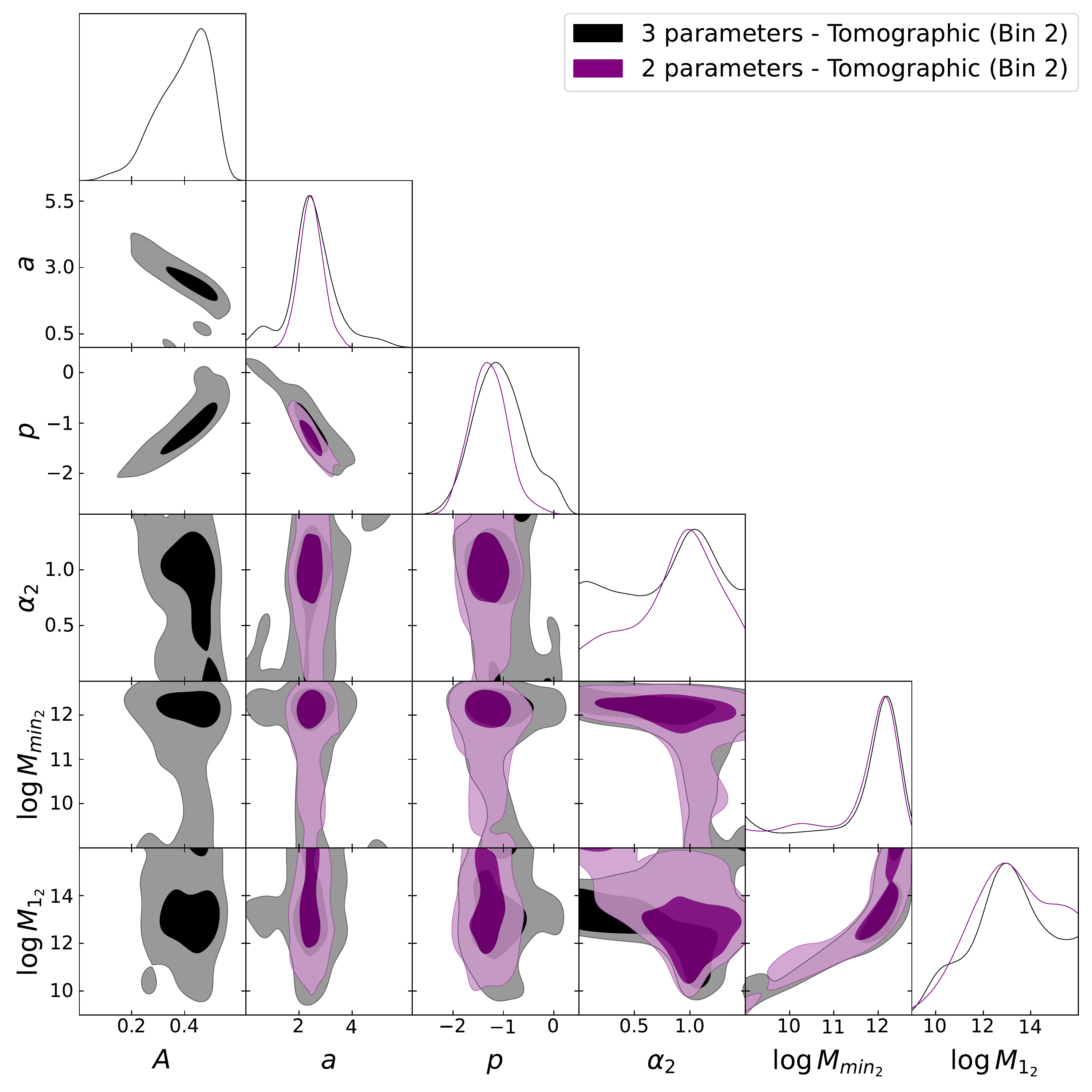} 
  \end{minipage} 
  \begin{minipage}[b]{0.5\linewidth}
    \centering  \includegraphics[width=0.9\columnwidth]{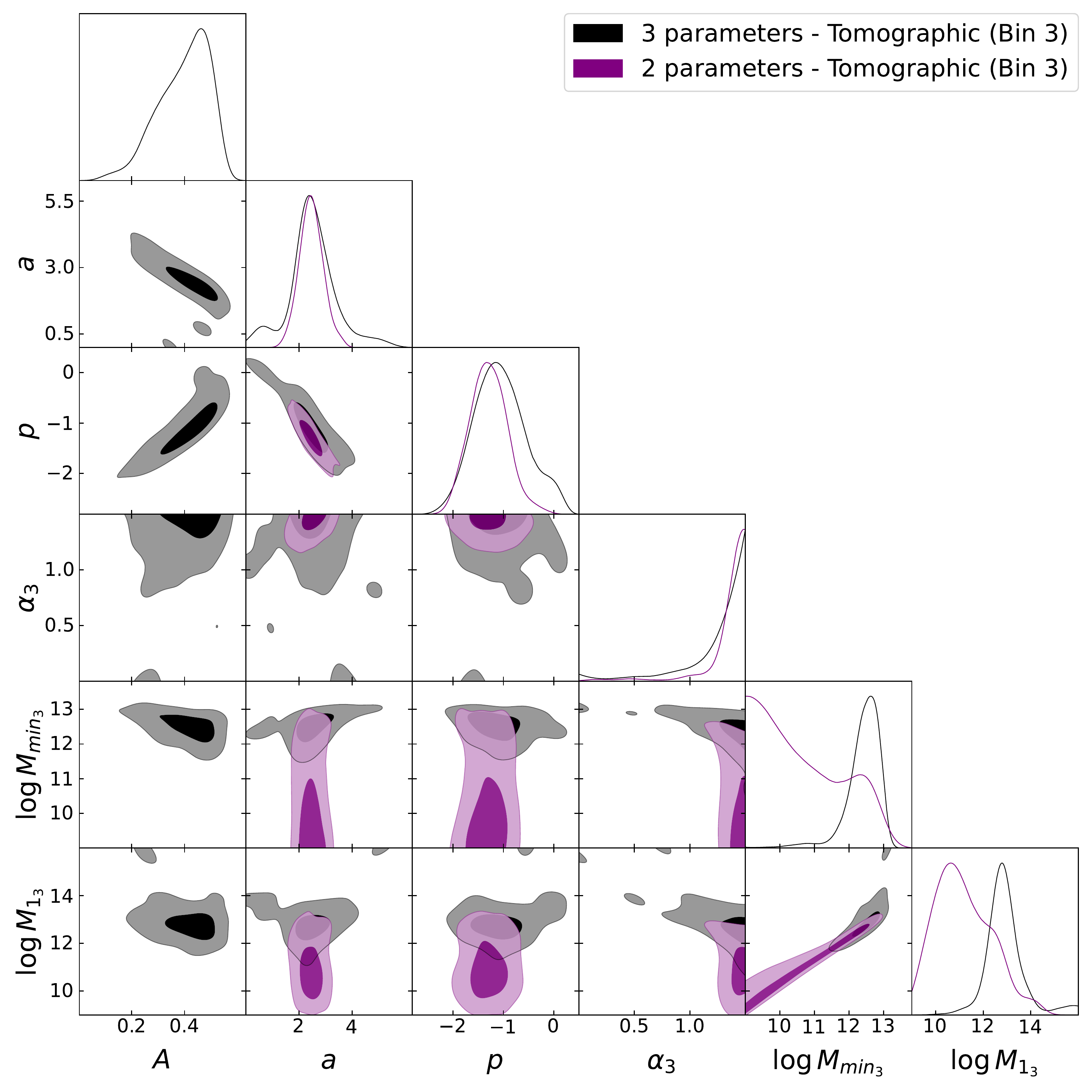}  
  \end{minipage}
  \begin{minipage}[b]{0.5\linewidth}
    \centering
    \includegraphics[width=0.9\columnwidth]{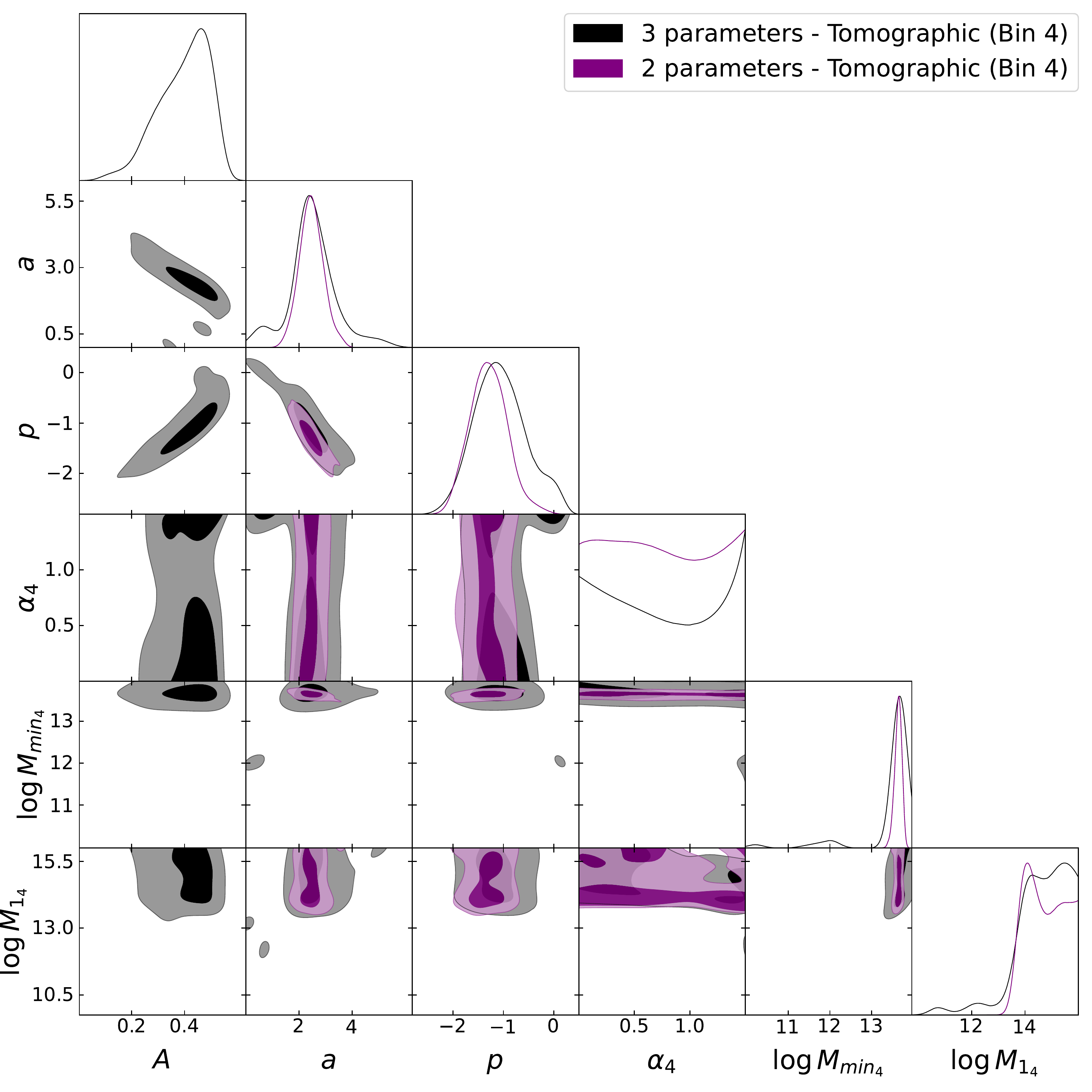}  
  \end{minipage} 
  \caption{One- and two-dimensional (contour) posterior distributions for the HMF and HOD parameters from the tomographic runs in the minitiles scheme for the three-parameter (in black) and the two-parameter (in purple) cases. The results from bins 1 to 4 are shown, respectively, from left to right and from top to bottom.}
  \label{corner_tomo_3param_vs_2param}
\end{figure*}

\begin{figure*}[h] 
  \begin{minipage}[b]{0.5\linewidth}
    \centering
    \includegraphics[width=0.9\columnwidth]{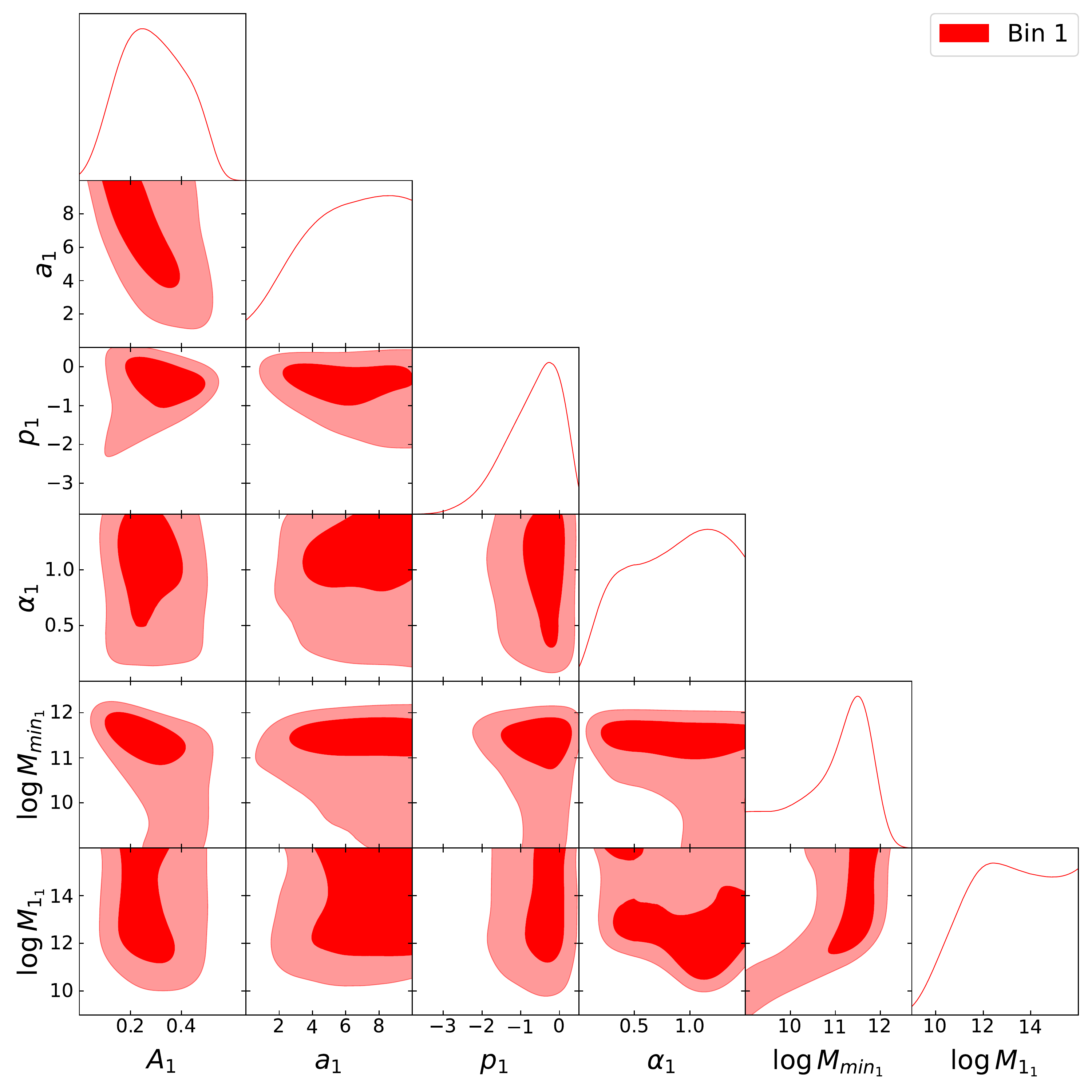} 
  \end{minipage}
  \begin{minipage}[b]{0.5\linewidth}
    \centering
    \includegraphics[width=0.9\columnwidth]{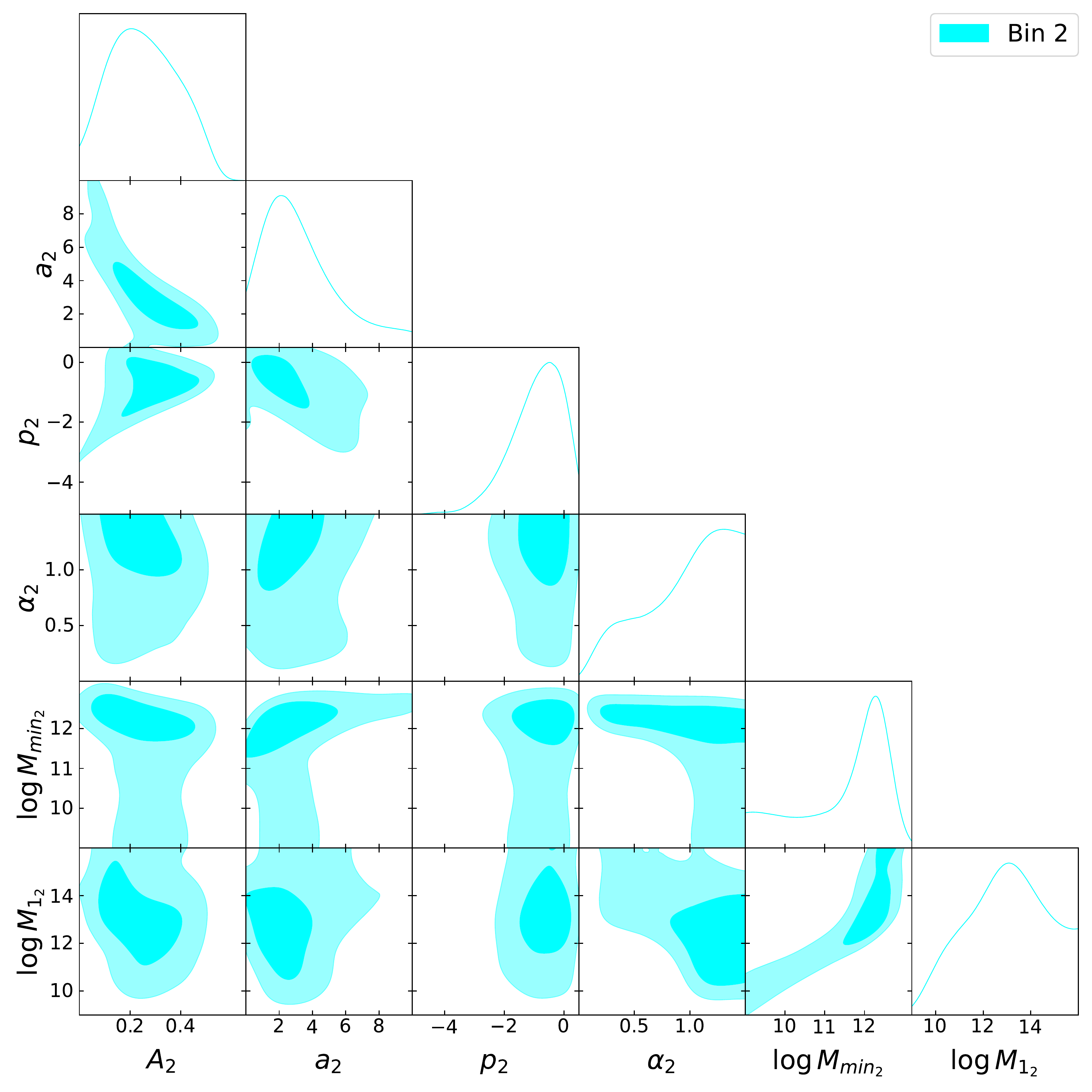}  
  \end{minipage} 
  \begin{minipage}[b]{0.5\linewidth}
    \centering
  \includegraphics[width=0.9\columnwidth]{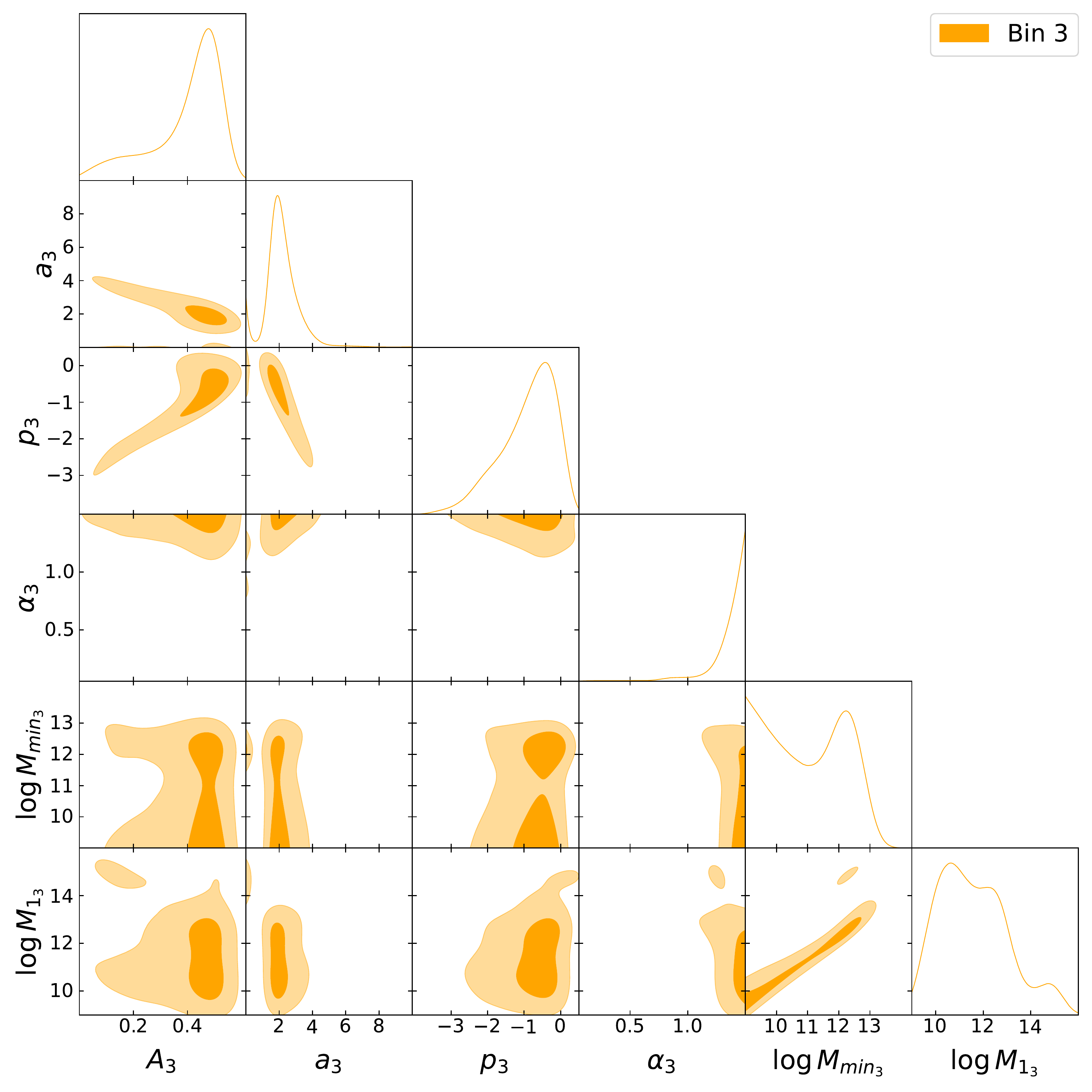}  
  \end{minipage}
  \begin{minipage}[b]{0.5\linewidth}
    \centering
    \includegraphics[width=0.9\columnwidth]{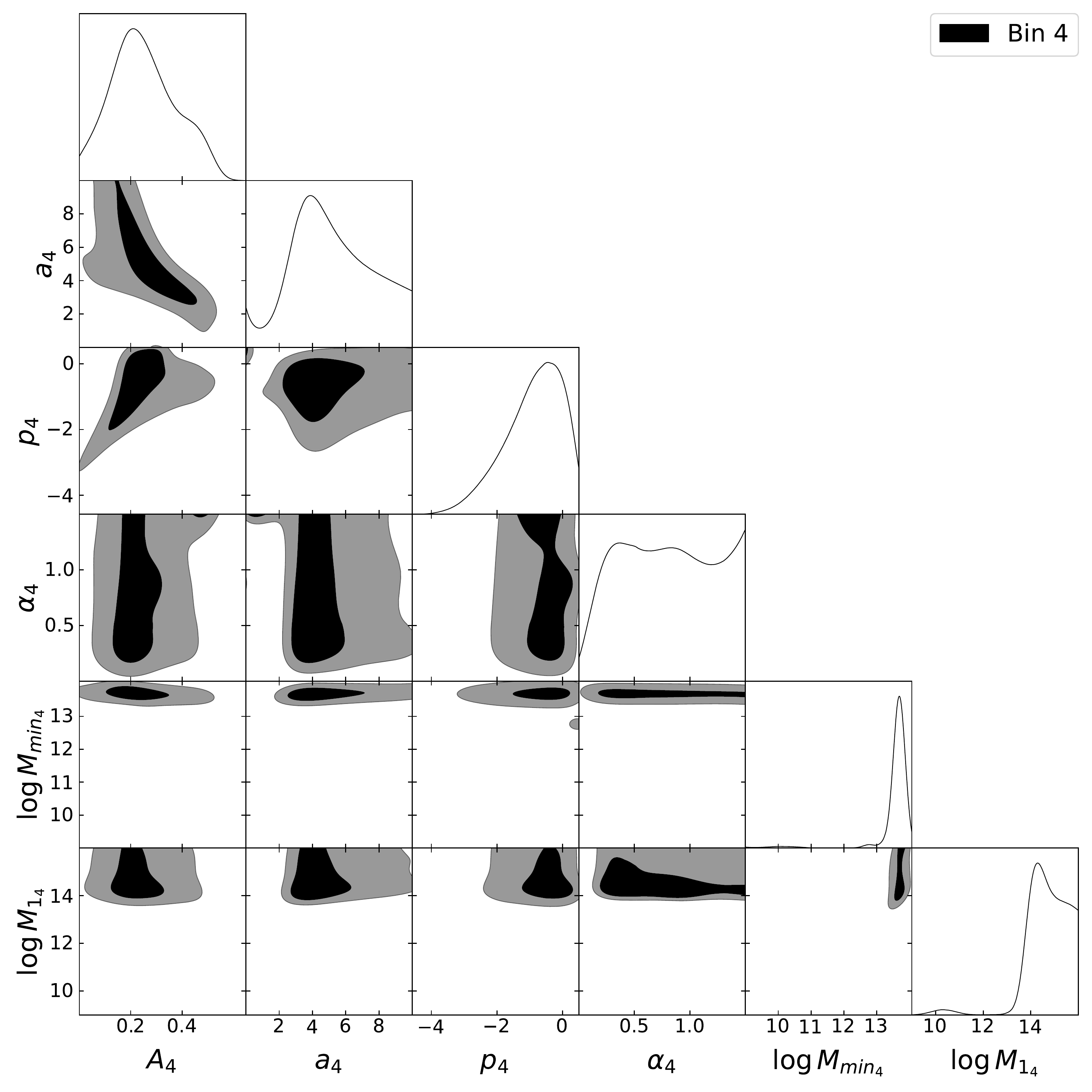} 
  \end{minipage} 
  \caption{One- and two-dimensional (contour) posterior distributions for the HMF and HOD parameters from the last four non-tomographic runs. The results from bins 1 to 4 are shown, respectively, from left to right and from top to bottom.}
  \label{tomo_caseB_cornerplot}
\end{figure*}

\end{document}